\documentclass[11pt,a4paper]{article}
\pdfoutput=1
\usepackage{jinstpub}
\usepackage{graphics}
\usepackage[caption=false]{subfig}
\usepackage{tabularx}
\usepackage{dcolumn}
\usepackage[sort&compress,numbers]{natbib}
\newcommand{\ctfe}{C$_3$F$_8$}
\newcommand{\degc}{$^{\circ}$C}

\title{A buffer-free concept bubble chamber for PICO dark matter searches}

\author[a,1]{M. Bressler}
\emailAdd{mjb536@drexel.edu}
\note[1]{Corresponding Author}
\author[a]{P. Campion}
\author[a]{V.S. Cushman}
\author[a]{A. Morrese}
\author[a]{J.M. Wagner}
\author[a]{S. Zerbo}
\author[a]{R. Neilson}
\affiliation[a]{Department of Physics\\
Drexel University; Philadelphia, PA 19104, USA}
\author[b]{M. Crisler}
\affiliation[b]{Particle Physics Division\\
Fermi National Accelerator Laboratory; Batavia, IL 60510, USA}
\author[c]{C.E. Dahl}
\affiliation[c]{Department of Physics and Astronomy\\
Northwestern University; Evanston, IL 60208, USA}

\date{\today}

\abstract{

In this paper, we report on the successful operation at Drexel University of the PICO collaboration's first \ctfe\ buffer-free prototype fluorocarbon bubble chamber. Previous PICO bubble chambers have produced world-leading WIMP search results with fluorocarbon target fluids, separated from the steel bellows by a buffer layer of water. Surface tension effects at the jar walls and liquid-liquid interface lead to a class of background events which have nuclear-recoil-like acoustic signatures and thus contaminate the WIMP-like signal region in analysis. Thus new bubble chambers are to be constructed ``right-side-up'', meaning that the jar of \ctfe\ is above the bellows with no water inside the inner vessel. The Drexel Bubble Chamber (DBC), runs successfully at and below the nuclear recoil thresholds used by PICO for WIMP searches, including thresholds as low as 1.19~keV. We have demonstrated sensitivity to $^{137}$Cs gammas and spontaneous fission neutrons from $^{244}$Cm, and acoustic alpha discrimination is demonstrated to be possible although the observed rate of alpha decays is very low. Position reconstruction from stereoscopic imaging allows for basic analysis cuts for calibration data. The successful construction and operation of this prototype confirm the effectiveness of the right-side-up design, which will be used in future dark matter searches such as PICO-40L and PICO-500.}

\begin{document}
\maketitle
\flushbottom

\section{Introduction}
Bubble chambers were invented by Donald Glaser in 1952 with the intent of being to elementary particle physics what the cloud chamber was to traditional nuclear physics \cite{glaser}. Bubble chambers work by maintaining a liquid below its vapor pressure, so that it is superheated, and waiting for a particle interaction to nucleate one or more bubbles to initiate boiling. As the boiling starts, cameras take and save photographs of the event and the bubble chamber is pressurized (the ``compression'') to a pressure above the vapor pressure to recondense the fluid, and then the pressure can be reduced again (the ``expansion'') to low pressure to reactivate the chamber. The initial growth of a bubble produces an audible sound, which was used to trigger the cameras on early bubble chambers \cite{characteristics} and can also be analyzed for particle identification. Thus the two basic forms of raw data from a bubble chamber event are the camera images and the acoustic response. This paper describes a new bubble chamber constructed at Drexel University as a prototype for future dark matter search experiments by the PICO collaboration.

\subsection{Bubble Nucleation}
The physics of bubble nucleation by particle interactions in bubble chambers was described by Seitz in 1958 \cite{Seitz}. Seitz' model is known as the ``hot spike'' model because it requires that a particle deposit a threshold amount of energy, $E_t$, inside of a small region creating hot spikes which ``literally explode into bubbles of larger than critical size'' (Seitz' words in \cite{Seitz}). The critical size for a bubble is a sphere of radius $R_c$, given by:
\begin{equation}\label{Rc}
R_c=\frac{2\sigma}{P_v-P_l}
\end{equation}
such that bubbles of radius $R>R_c$ will expand and grow to macroscopic size, while bubbles with $R<R_c$ will collapse due to surface tension. The denominator of equation \ref{Rc} is the difference between the pressure inside the bubble (assumed to be the vapor pressure at the given temperature) and the pressure of the liquid (set by the apparatus), and $\sigma$ is the liquid's surface tension.

The threshold energy deposition to create a bubble of at least critical size, as presented by the COUPP collaboration in 2013 \cite{COUPPCF3I} (a full derivation is available in appendix A of \cite{ERpaper}) is given by 
\begin{equation}\label{Et}
E_t=4\pi R_c^2(\sigma-T(\frac{\partial \sigma}{\partial T})_\mu)+\frac{4\pi}{3}R_c^3\rho_v(h_v-h_l)-\frac{4\pi}{3}R_c^3(P_v-P_l)
\end{equation}
where subscript ``$v$'' means a value for the vapor inside of the bubble, while subscript ``$l$'' means a value for the surrounding liquid, $\rho$ is density, $T$ is the temperature, and $h$ is specific enthalpy. The subscript $\mu$ in $(\frac{\partial \sigma}{\partial T})_{\mu}$ indicates that the vapor and liquid are in chemical equilibrium and the rate of pressure rise is constant. This is the expression used by PICO to determine thresholds, and includes a surface term representing energy stored in surface tension (first term), a volume term indicating the energy needed to boil a small sphere of liquid (second term), and a term accounting for reversible work done in expanding the bubble (third term). This expression is similar to that derived by Seitz in \cite{Seitz}, with  corrections for the temperature dependence of the surface tension in the surface-energy term, and the addition of the reversible-work term. The threshold energy must be deposited in the region of liquid that will eventually grow to the size of the critical bubble.

\subsection{PICO Bubble Chambers}
Traditionally, bubble chambers such as the Big European Bubble Chamber (BEBC) at CERN were used as tracking detectors in high-energy collider experiments \cite{BEBC}. These high-energy-physics chambers' cameras recorded the images of trails of bubbles that ionizing particles left as they passed through the detectors. The chambers were equipped with large magnetic fields so that positively- and negatively-charged particles would spiral in opposite directions, and cycles of expansion and compression were rapid and synced with the arrival of accelerator beams, so that the superheat would only need to be maintained for a fraction of a second \cite{bebctechnote}. By contrast, PICO's dark matter search bubble chambers need to maintain the superheated state for long periods of time in order to maximize WIMP search exposures, and are operated at pressure/temperature combinations that make the detector insensitive to minimum-ionizing radiation. PICO bubble chambers instead seek nuclear-recoil events, as would be produced by a WIMP-nucleon interaction \cite{WIMPdetection}, but also by scattering of environmental neutrons. Whereas in a tracking detector, the particle's signature is the tightness and direction of its spiral in a track of thousands of bubbles, in a dark matter search bubble chamber a neutral particle initiating a nuclear recoil leaves only a single bubble at its interaction point. Detailed neutron calibration experiments performed by PICO find that the bubble nucleation efficiency for fluorine recoils rises from 0 to 100\% slightly above the threshold energy calculated from Equation \ref{Et}. There is no track information in PICO bubble chambers, except that a neutron can multiple-scatter in the detector and create multiple bubbles, but with space between them and no indication of which one was produced first. 

PICO's WIMP search bubble chambers are operated deep underground for shielding from cosmic radiation, and materials are selected based on known or measured radiopurity. Detailed simulations are performed to determine expected background rates. Piezoelectric transducers attached to the glass jars of the bubble chambers record the sound of bubble nucleation, which is used to discriminate alpha decays from nuclear recoils with a parameter called ``AP'', roughly the acoustic power carried by the sound \cite{alphadiscrimination}; low AP corresponds to nuclear recoils while high AP corresponds to alpha decay of dissolved radon and its progeny. The phenomenon of higher-AP alpha decay nucleation is attributable to the ability of alphas to deposit large amounts of energy over relatively long distances, as has been modeled with molecular dynamics in \cite{alphasMD}. These bubble chambers have set world-leading limits on the spin-dependent WIMP-proton coupling with fluorocarbon target fluids \cite{pico60c3f8,2Lrun1,2Lrun2}.

PICO chambers have historically included a buffer layer of ultra-pure water to isolate the superheated target fluid from all surfaces except the synthetic silica jars, as described in \cite{2Lrun1,COUPP4,60cf3i}. This water layer was found to be a contributing factor to a class of background events which mimick a WIMP signal in acoustic characteristics \cite{2Lrun2}, which were caused by surface tension effects of water droplets attached to particulate from the glass and steel. These background events were not uniformly distributed in time and space as they would be for a true WIMP signal, and a combination of timing, spatial and acoustic cuts has proven partially effective in discriminating against this class of background \cite{60cf3i,2Lrun1}.

An obvious long-term solution to removing this background is elimination of the need for water in contact with the active fluid. In the design presented in this paper, the only liquid in the inner volume of the detector is the target fluorocarbon, similar to a propane bubble chamber operated in the 1960s by Waters, Petroff, and Koski for tritium and carbon-14 counting \cite{waterschamber}. This design will be utilized by the future WIMP search bubble chambers PICO-40L and PICO-500. PICO-40L is a 40-liter bubble chamber at SNOLAB \cite{snolab_site} in Sudbury, Canada, which will take WIMP search data starting in 2019, and PICO-500 is a ton-scale bubble chamber currently under design to be constructed at SNOLAB. 

Another small buffer-fluid-free bubble chamber of a different style, with liquid xenon target fluid, has already been operated at Northwestern University \cite{xebc}. The liquid noble target fluid allows for simultaneous collection of scintillation signals and bubble chamber operation, for increased particle identification capabilities and event-by-event energy reconstruction. Work is underway to scale up the scintillating bubble chamber technology with argon target fluid for use in WIMP and CE$\nu$NS searches \cite{ArBC}.

\section{Description of the Drexel Bubble Chamber}
\begin{figure}
\centering
\includegraphics[scale=0.4]{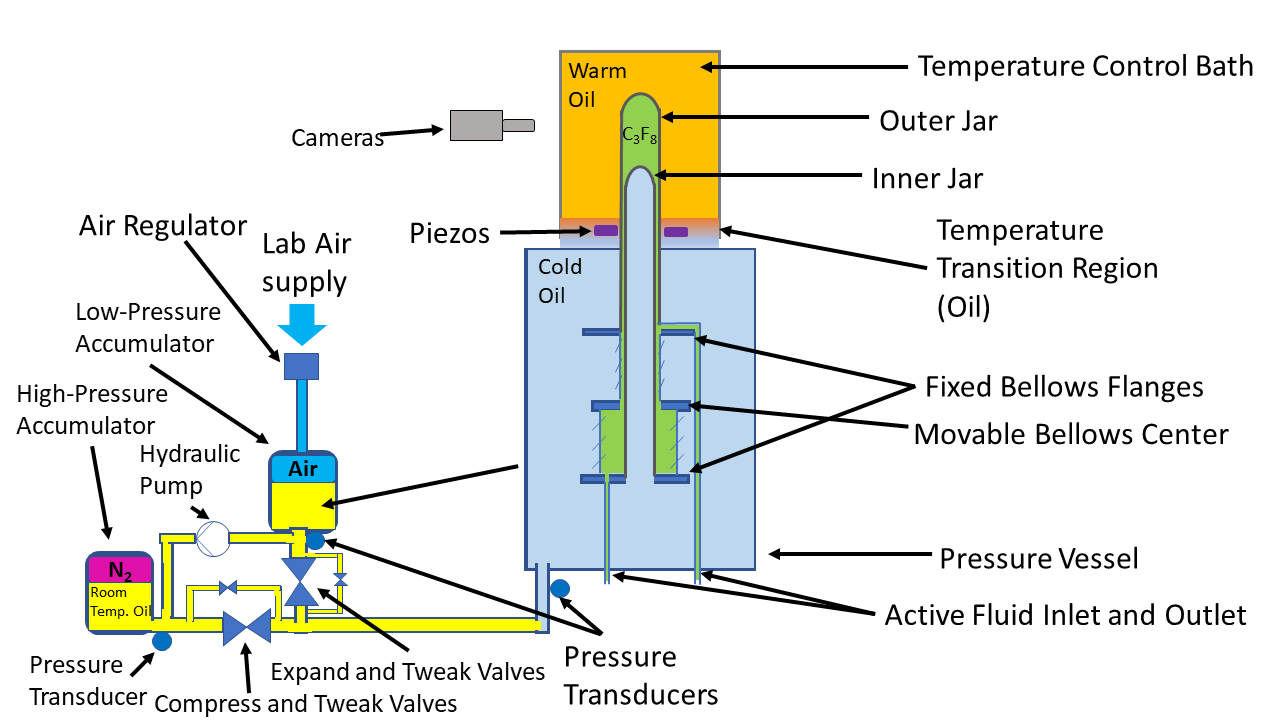}
\caption{\label{DBCschematic} A simplified schematic of the bubble chamber and hydraulic system, showing the bubble chamber's inner and outer jars, the bellows system, the heat bath, the placement of basic instrumentation, and the hydraulic system. Manual valves for filling and servicing the system are not shown. The pressurized \ctfe\ volume is partially contained within the pressure vessel, but extends up from the pressure vessel into a heat bath at ambient pressure. A thermal gradient separates the active superheated \ctfe\ in the heat bath from the stable \ctfe\ in the bellows region below.}
\end{figure}

\begin{figure}
\centering
\includegraphics[scale=0.4]{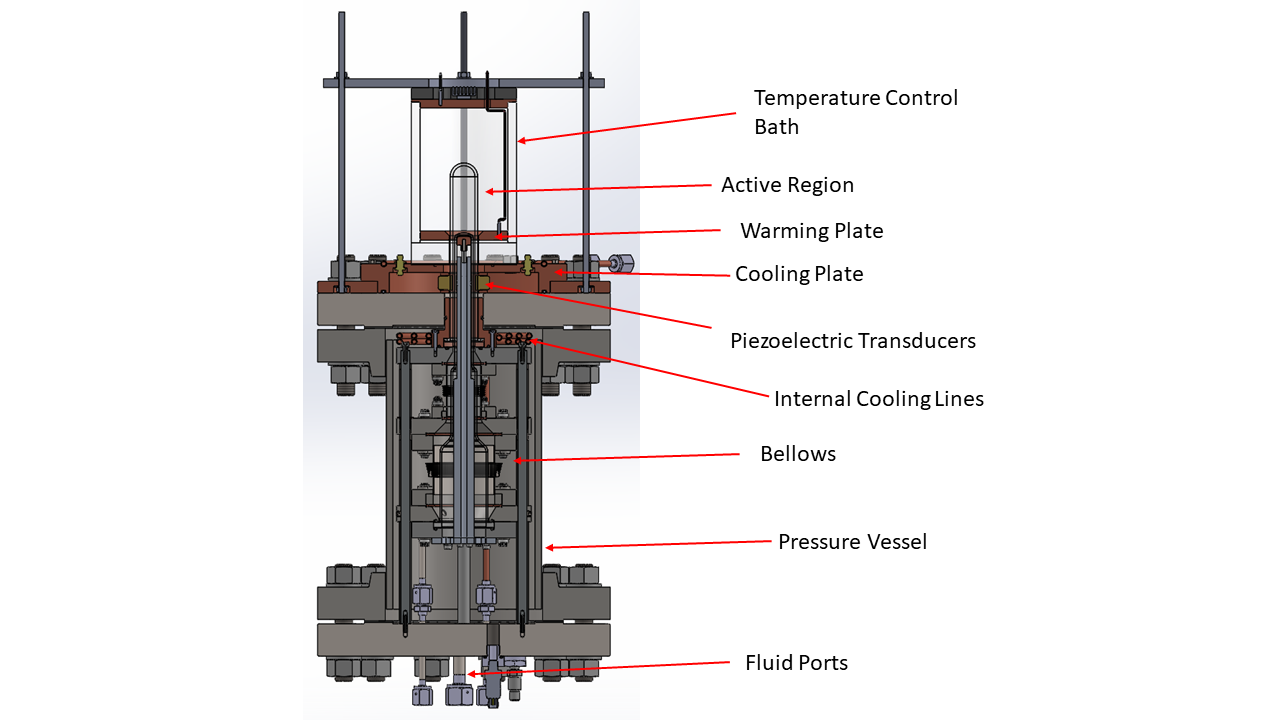}
\caption{\label{solidmodel} A 3D rendering of a cutaway of the DBC is shown, including the inner assembly, the pressure vessel, and the temperature control bath.}
\end{figure}

The schematic in Figure \ref{DBCschematic} shows the Drexel Bubble Chamber (DBC), which demonstrates proof-of-concept of the single-fluid bufferless design for fluorocarbon bubble chambers. To isolate the superheated liquid from the steel parts and gasket seals, a sharp temperature gradient is established; the region near the steel bellows is kept cold (ideally below -25\degc\ for \ctfe\ so that the fluorocarbon is not superheated during the expansion), while the glass jar in view of the cameras is kept warmer (\ctfe\ PICO bubble chambers typically operate between 14\degc\ and 22\degc). To prevent convection between the warm and cold regions, the warm region sits above the cold region, leading to the ``right-side-up'' name, indicating that the detector design is conceptually flipped compared to previous PICO bubble chambers. The inner vessel, pressure vessel, and temperature control bath are illustrated in figure \ref{solidmodel}. Easy access to the target fluid is provided by valves on the bottom of the apparatus, allowing for circulation and purification, which was prevented by the water buffer layer on previous PICO bubble chambers. The DBC was not built with radiopure materials as a PICO dark matter search detector would be, so backgrounds are expected from the materials themselves in addition to cosmic and environmental radiation.

\subsection{Inner Assembly}
\begin{figure}
\centering
\includegraphics[scale=0.08]{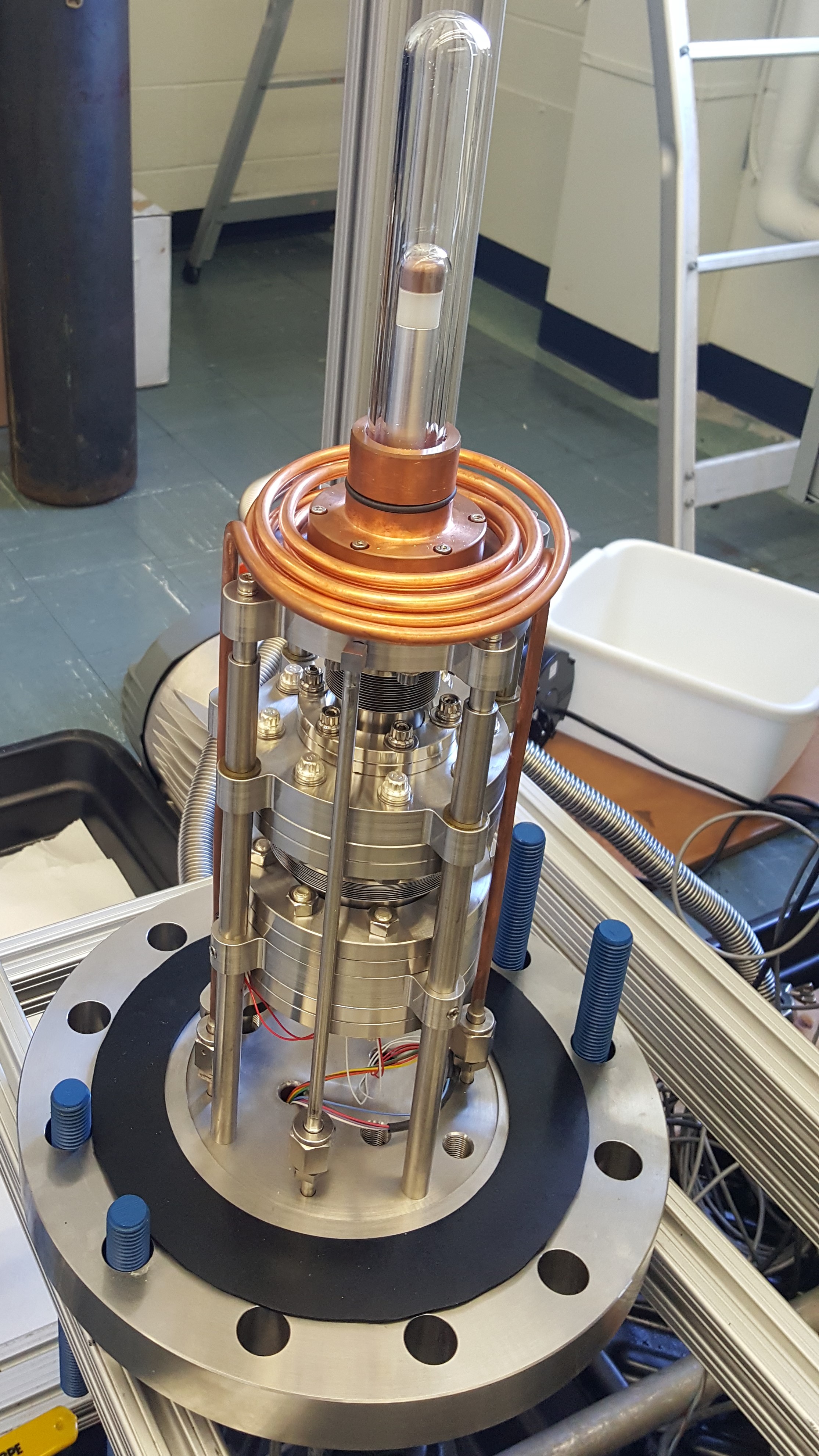}
\caption{\label{inner_assembly_photo} A photograph of the inner assembly prior to placement of the pressure vessel, showing the jars and bellows, mounted to the bottom flange of the pressure vessel. The copper cooling coil around the assembly is internal to the pressure vessel, as described in Section \protect{\ref{therm}}.}
\end{figure}

The inner assembly of the DBC, the design for which is shown in Figure \ref{solidmodel}, consists of a set of two fused-quartz glass test-tubes (the ``inner'' and ``outer'' jars) sealed to the bellows (described in Section \ref{Pcontrol}), with the \ctfe\ contained between them. The outer jar is a 24mm inner-diameter and 3~mm thick glass test-tube, the top $\sim$10~cm of which is contained within a temperature-control bath, filled with mineral oil (Petrochem white oil FG-10), which has operated at temperatures between 15\degc\ and 25\degc; this warm section defines the active region of the detector, and contains about 40~mL of \ctfe. The bottom of the outer jar descends into a steel pressure vessel which is kept cold, with target temperature -15\degc, so \ctfe\ in this lower part of the system is minimally (or not at all) superheated at the operational pressures, around 30~psia, and the jar is sealed to the stationary top of the bellows assembly. A smaller-diameter glass tube, the ``inner jar'', is fitted inside the outer jar to reduce the fluid volume outside of the active region, limit vertical motion of the fluid, and discourage heat conduction between the active fluid and the cold region. The inner jar is sealed to the stationary bottom of the bellows assembly. See Figures \ref{DBCschematic} and \ref{solidmodel} for schematic drawings of the setup, and Figure \ref{inner_assembly_photo} for a photo of the inner assembly before the pressure vessel was placed around it.

The \ctfe\ volume is filled, sealed, and drained through two bellows-sealed valves on the bottom flange of the cold pressure vessel, and its pressure is measured directly on the inner-volume side of one of these valves by a high-purity pressure transducer (Setra model number 225). Filling the inner volume with \ctfe\ is accomplished by cooling the entire assembly (including the normally-warm region) to $\sim$0\degc, connecting a room-temperature tank of \ctfe\ to the fill valve via stainless steel and copper tubes, and allowing the fluorocarbon to distill into the cold detector.

The glass jars are sealed to a 2-part stainless steel pressure-balancing edge-welded bellows assembly, the center of which is allowed to move vertically with a range of motion of 1.1~cm. The top half of the bellows assembly has a smaller diameter than the bottom half, so motion of the center upward increases the inner \ctfe\ volume, reducing the pressure, while motion downward increases the pressure; thus the hydraulic oil pressure inside the pressure vessel directly determines the pressure of the \ctfe. The hydraulic oil is the same brand and type as the oil contained in the temperature-control bath.

\subsection{Pressure Control and Hydraulic System}
\label{Pcontrol}
A simplified schematic of the DBC's hydraulic system is shown in Figure \ref{DBCschematic}. In the compressed state, the bubble chamber is kept around 200psia (oil pressures read by transducers made by Omega, model number PX319-300A5V) by a high-pressure bladder accumulator (Greer model number BA01B3T01A1G) charged with nitrogen gas. When ready for an expansion, the hydraulic system closes the ``compress'' valve, and opens the ``expand'' valve leading to a low-pressure accumulator (Greer model number BA002B3T01A1G), which brings the pressure in the pressure vessel/bubble chamber system down to the set expand pressure, typically between 20 and 60~psia. The system regulates the pressure during the expanded phase with air pressure via a PID-controlled air regulator (a PROPORTION-AIR QPV1). An alternative, less precise pressure regulation scheme is available when there is no supply of compressed air, in which ``tweak'' valves move small amounts of hydraulic oil in and out of the pressure vessel to regulate the pressure. When a trigger is initiated as described in Section \ref{triggering}, the hydraulic system recompresses the bubble chamber to $\sim$200~psia by closing the expand valve and opening the compress valve, quenching any boiling. After 30~s, the system expands again, starting the next cycle. After every 30 events the system is held in the compressed state for an extended time of 300~s. The net effect of the continuous expand-compress-expand cycling is to transfer oil from the high-pressure to the low-pressure accumulator, thus a hydraulic pump (a piston pump made by Clark Solutions, model ET-150) returns oil to the high-pressure accumulator during the compressed time. Because the pressure transducer used for pressure control is placed 65~cm below the active region, the pressure $P_{\text{expand}}$ must be corrected by subtracting 1.3~psi from the pressure transducer value to get the pressure $P_{\text{chamber}}$ in the active region, to account for the hydrostatic pressure difference. 

The position of the movable center of the bellows is read out by a Gefran magnetostrictive linear position sensor, which is sensitive to displacements of order 0.1~mm. A resistance temperature detector (RTD) on the center piece of the bellows measures the cold region temperature, an RTD inside the inner jar measures the temperature gradient between the cold and warm regions, and another Omega pressure transducer measures the hydraulic oil pressure in the pressure vessel.

\subsection{Thermal Systems}
\label{therm}
The cold region is kept cool with PolyScience circulating baths (model numbers AD07R-40-A11B and AD15R-40-A11B) which pump PolyScience HC-50 cooling fluid through copper tubes internal to the pressure vessel, which is covered in foam insulation. Some cooling fluid is also routed to tubes on select external components, such as the top and bottom flanges of the pressure vessel, and the active fluid inlet and outlet ports.

The active region is kept warm with a convective temperature-control bath of mineral oil contained by an acrylic vessel. The oil is warmed by a resistive heating element on a copper plate at the bottom of the acrylic vessel, and convection is achieved by cooling another copper plate at the top of the vessel with a peltier element (CUI Inc. part number CP85338). An RTD is embedded in the copper plate at the bottom, and the oil temperature is measured by an RTD suspended in the oil near the bubble chamber.  The RTDs' readouts are used as feedback for two BK Precision 1687B PID-controlled switching-mode power supplys which control the heating element and  the peltier.

\subsection{Cameras and Acoustic Sensors}
\begin{figure}[h!]
\centering
\includegraphics[scale=0.08,angle=-90]{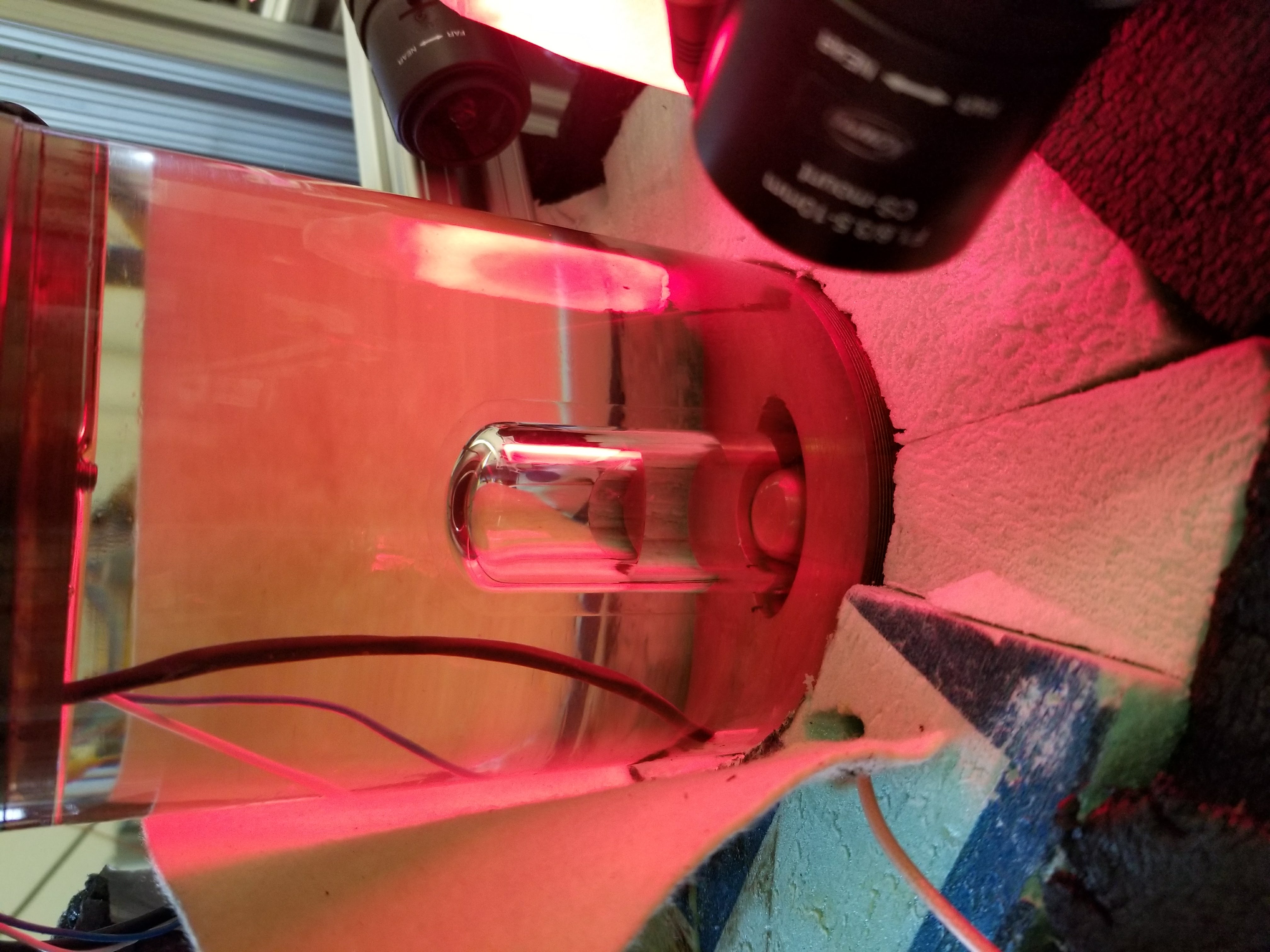}
\caption{\label{camphoto} A photograph of the active part of the bubble chamber inside the temperature-control bath. The cameras view at right angles to capture perpendicular views of bubbles for simple position reconstruction. This image was taken while filling the chamber with \ctfe; the surface of the liquid is visible about halfway up the visible part of the bubble chamber.}
\end{figure}

Two Basler cameras (Model acA640-120gm) external to the temperature-control bath view the chamber at right angles to each other to capture a stereoscopic view of bubble formation, as shown in the photo in Figure \ref{camphoto}. The cameras normally operate at 100~Hz and act as the primary trigger. The camera images are saved around the time of the trigger; during normal operation the 5 frames before the trigger and the 5 frames after the trigger are recorded with the event's raw data.

Four piezoelectric transducers (piezos) custom-made by PICO are pressed against the outer jar to record the sound of bubble nucleation between the pressure vessel and the heat bath. The piezos' placement against the outer jar is shown in Figure \ref{piezos}. For the data presented below, only two of the four piezos (numbers 1 and 4) are read out and analyzed. The piezo signals are individually amplified by custom-made amplifier boxes before entering the data acquisition system.

\begin{figure}[h!]
\centering
\includegraphics[scale=0.08,angle=-180]{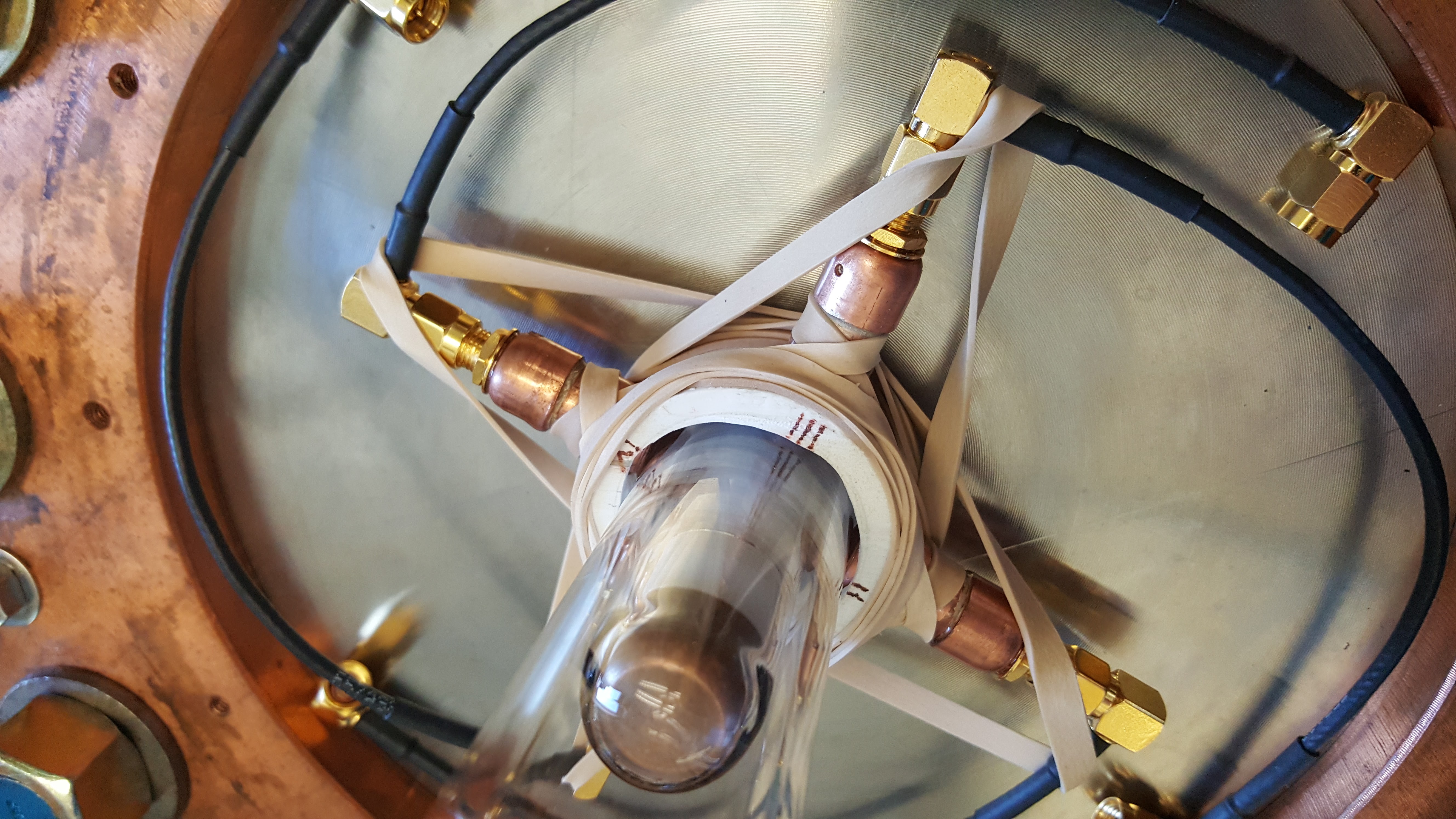}
\caption{\label{piezos} A photograph of the placement of piezoelectric transducers against the glass of the outer jar of the bubble chamber. The piezos are pressed against the glass with rubber bands for ease of assembly and disassembly, and held in their locations with a PVC ring. The piezos are located below a copper cooling plate (not shown), and are not accessible or visible during normal operation.}
\end{figure}

\subsection{Controls and Data Acquisition}
The DBC is controlled by a National Instruments CompactRIO (NI and cRIO for short), which communicates with the data acquisition systems and storage via a LabVIEW VI (virtual Instrument), written by PICO for general bubble chamber operations, run on a Windows 7 PC. The VI displays diagnostic information to the bubble chamber operator in real time, so the system can be monitored and controlled continuously.

The piezos are read out through a NI TB-2708 simultaneous-sampling connector and NI PXI-6133 multifunction I/O module into a NI PXI-1044 Chassis, which also handles trigger and camera timing with a connection to the PC and camera trigger line through a NI USB-6501 digital I/O module. The camera data is sent directly to a network adapter installed on the control PC via Ethernet.

Slow readout instruments, including the pressure transducers, the RTDs, and the bellows position sensor are readout through analog input modules (NI-9219, NI-9201, and NI-9217) on the cRIO. The cRIO also controls the hydraulic system valves, pump, and air regulator through analog output modules (NI-9266 and NI-9263). 

\label{triggering}
There are 4 configurable classes of triggers that can occur to end an expansion: camera, pressure, timeout, and manual. A camera trigger is initiated when either camera detects that a set number of pixels have changed from one frame to the next, a pressure trigger is initiated when one or more of the pressure transducers detect a pressure change that could indicate boiling, a timeout trigger is initiated if the chamber has stayed in the expanded state for a pre-determined amount of time (for the data presented here this is 300~s), and a manual trigger is when the bubble chamber operator presses the ``manual trigger'' button on the DAQ. The trigger code initiating the compression is saved with the event's metadata. An event is defined by one cycle of
\begin{enumerate}
	\item{ 30 or 300 second compression}
	\item{ Expansion}
	\item{ Live expanded time}
	\item{ Trigger}
	\item{ Recompression.}
\end{enumerate}

\section{Data Analysis and Results}
The DBC saw first operation in June 2017 and has since shown responses to fast neutrons and gamma rays consistent with previous PICO bubble chambers, and has demonstrated the anticipated ability to remain superheated for a few seconds after the nucleation of a bubble. The data presented here were all taken in an above-ground lab with little shielding, so most of the background is made up of fast neutrons from cosmic ray interactions.

\subsection{Optical Event Reconstruction}
\begin{figure}
\centering
\includegraphics[width=0.9\textwidth]{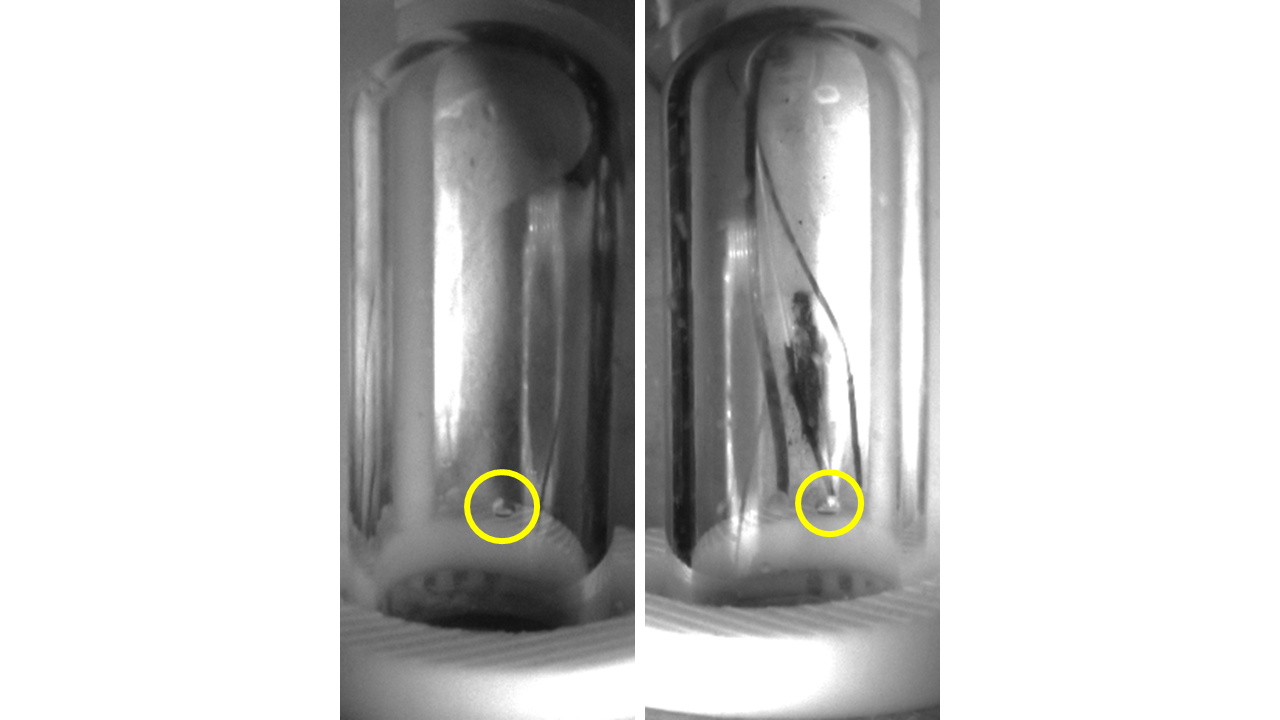}
\caption{\label{examplebubble} A typical single-bubble event in the DBC, as seen in each camera.}
\end{figure}

\begin{figure}
\centering
\includegraphics[width=0.9\textwidth]{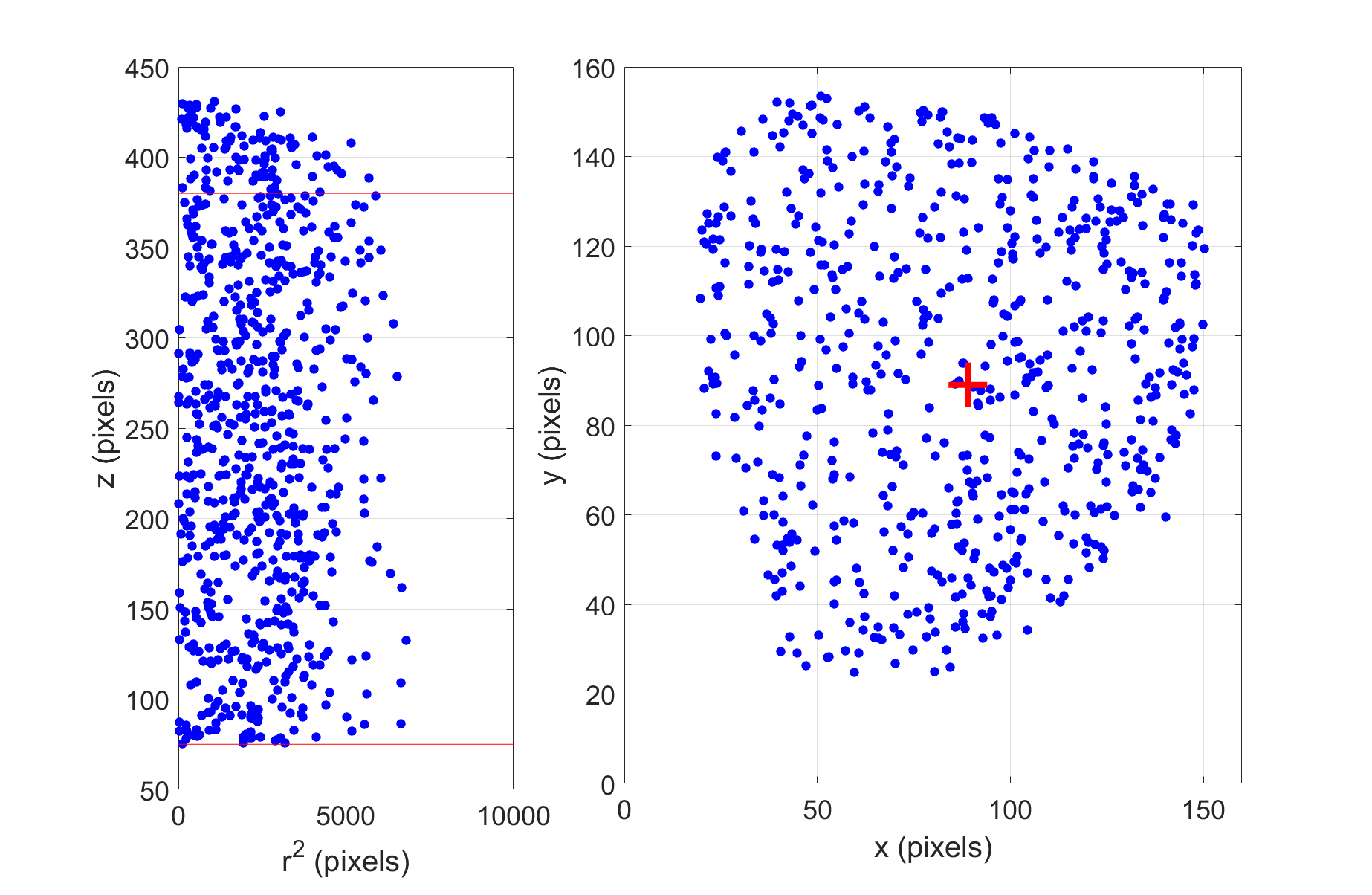}
\caption{\label{proj} Left: A projection of $r^2$ vs $z$ of the events from a background run taken at 35~psia expansion pressure and 20\degc; red lines indicate the z cut (lower) and the start of the dome region (upper). Right: The $x$-$y$ projection of events from the run, with the center of the chamber shown as a red cross. The projection is slightly out-of-round due to lensing effects not accounted for in this analysis. This projection does not include bubbles originating near the top of the chamber in the hemispherical ``dome''. }
\end{figure}

Position reconstruction of the bubbles from the saved images is done automatically by image analysis software developed specifically for this task by PICO. An example pair of images showing a typical single bubble are shown in figure \ref{examplebubble}. The bubble-finding code uses image subtraction to find clusters of pixels that have changed between consecutive frames, and saves the pixel coordinates of the cluster for the first frame in which the bubble becomes visible. The two cameras viewing the DBC are placed at 90$^\circ$ to each other, thus position reconstruction is relatively easy, with a natural definition of $x$ and $y$ being the horizontal position of the bubble in the image from each camera, so $r^2 = (x-x_{\text{center}})^2 + (y-y_{\text{center}})^2$, where $x_{\text{center}}$ and $y_{\text{center}}$ are the center pixels in the images from each camera, and $z$ is chosen as the vertical position of the point in one of the cameras. All coordinates are kept in ``pixel'' space for simplicity, and the lensing from the glass, oil, and acrylic is not corrected, leading to a guitar-pick-shaped $x$-$y$ projection, as shown on the right in Figure \ref{proj}.

\subsection{Event Selection}
Events for which no bubble is found by the bubble-finding software and events for which there is disagreement between the determined number of bubbles in images from the two cameras are cut from the analysis. A cut is chosen in the vertical ($z$) coordinate to exclude bubbles that nucleate out of the view of the cameras and rise into the visible volume. This also cuts out the region where the temperature gradient is established. Events with very short expands are removed, ensuring that for each event kept in the analysis, the chamber has reached its steady expanded pressure. For each data set the expand time cut is the mean time after the start of the expansion to when the chamber comes within 2~psia of the set point. For runs when the pressure is regulated by air pressure rather than by tweak valves, an additional 4 seconds are added to the calculated expand time cut to account for the tendency of the pressure to ``overshoot'' the setpoint (dropping below $P_{\text{expand}}$ by a few psi) before settling to the steady expanded state. Typical cuts are in the range 5-10 seconds, depending on the pressure set point. During operation without radioactive sources, the live fraction (live time divided by actual elapsed time) is typically above 50\%.

\begin{figure}
\centering
\includegraphics[width=\textwidth]{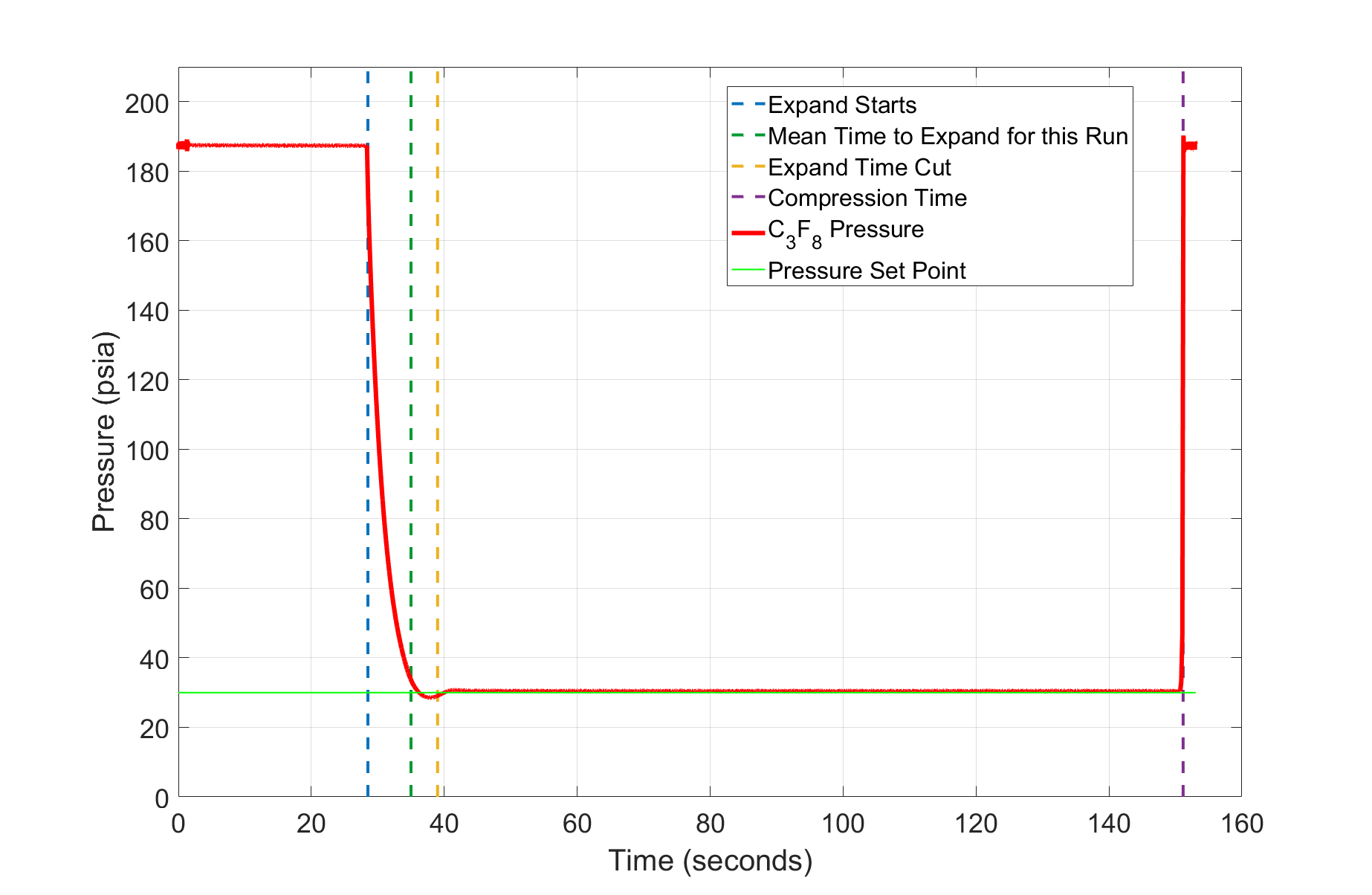}
\caption{\label{example_expand} An example expansion cycle in the DBC, with 112~seconds of live time between the expand time cut and the compression. The \ctfe\ pressure is shown as a thick red line, the pressure set point is shown as a thin green line, the blue dashed line marks the start of the expansion, the green dashed line marks the mean time it took to reach within 2~psia of the setpoint for all events in this run, the dashed orange line marks the expand time cut for this event, and the dashed purple line indicates the time of the compression. The ``overshoot'' is visible between the dashed green and orange lines.}
\end{figure}

\subsection{Response to neutrons}
Neutrons from neutron sources, are used to calibrate dark matter search bubble chambers and test chambers. A neutron incident on a target nucleus produces a similar nuclear recoil to that of a WIMP interaction, so neutrons play an important role in the background to the WIMP search and in detector calibration.

\begin{figure}[h!]
\centering
	\subfloat{\includegraphics[width=0.5\linewidth]{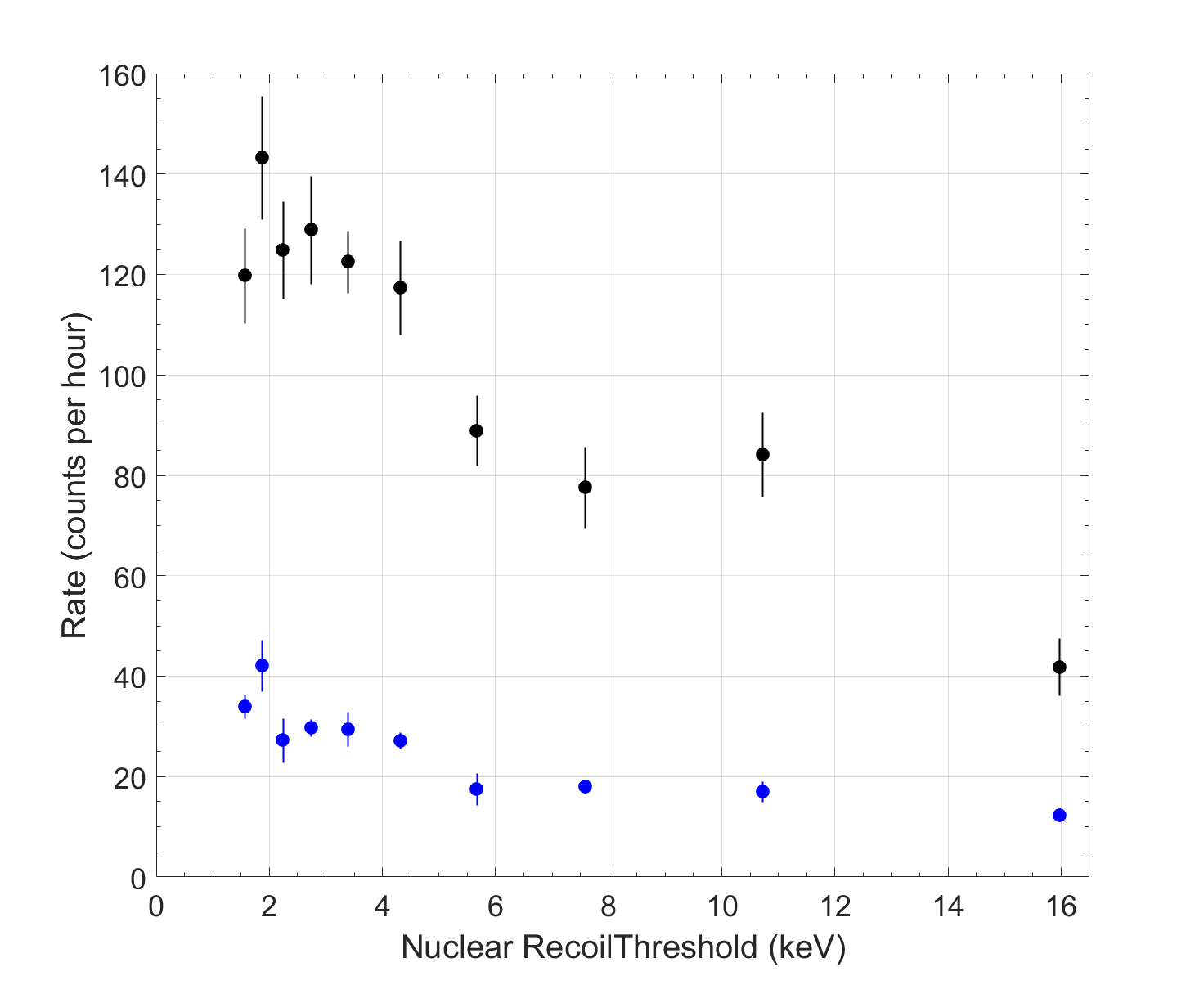}}%
	\subfloat{\includegraphics[width=0.5\linewidth]{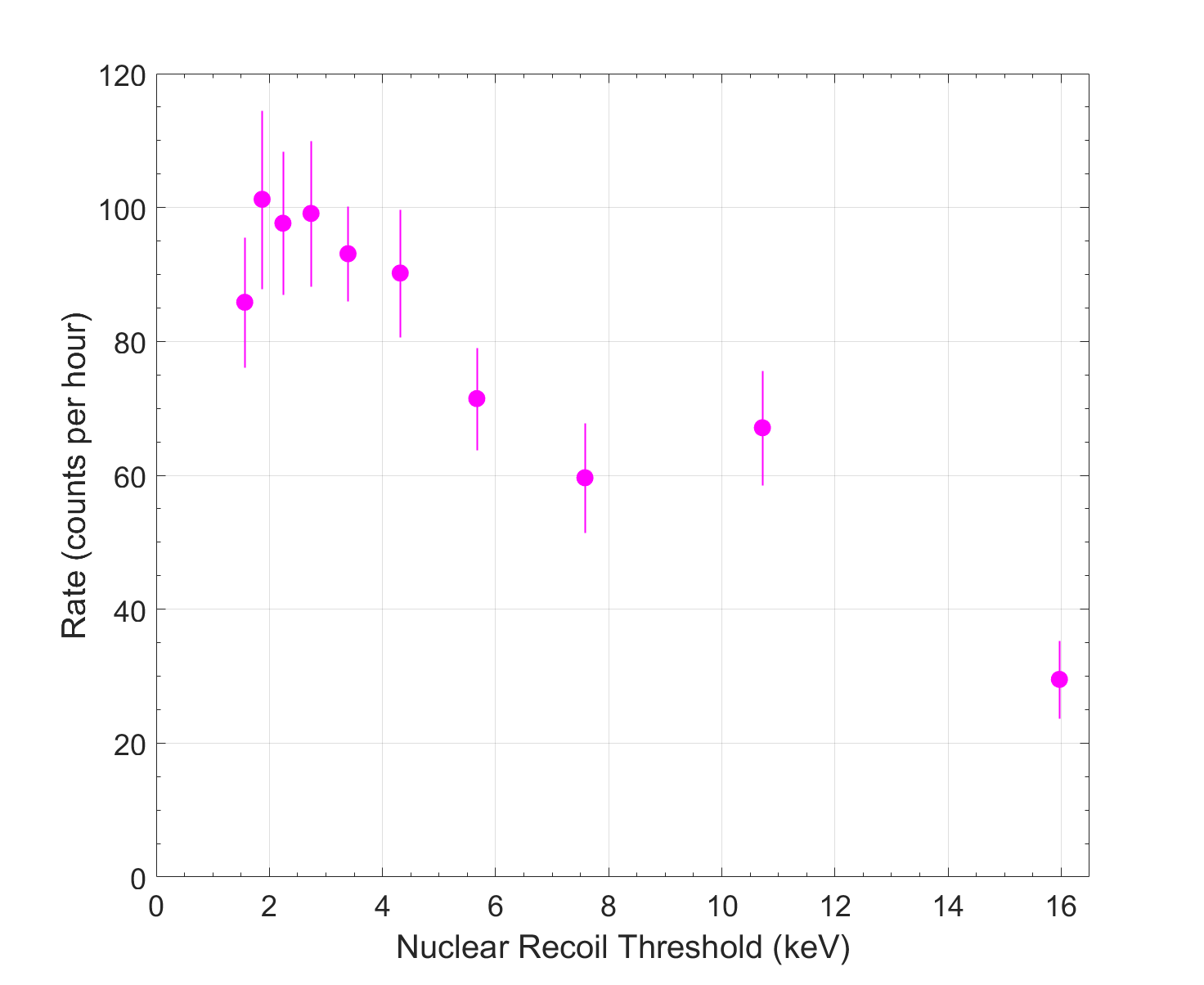}}
\caption{\label{neutronrate} Left: Raw rates with (black) and without (blue) the neutron source; Right: Rate of bubbles nucleated by neutrons from the $^{244}$Cm source after background subtraction. The response to neutrons shows a decreasing trend with increasing nuclear recoil threshold, indicating that the DBC is sensitive to recoils from the lower-energy neutrons at lower nuclear recoil thresholds.}
\end{figure}

A 34~$\mu$Ci $^{244}$Cm source was placed 10~cm above the DBC resting on the lid of the oil bath to provide fast neutrons from spontanaeous fission. $^{244}$Cm primarily undergoes alpha decay, but the alpha radiation is stopped quickly and does not make it into the bubble chamber. The branching ratio for spontaneous fission is 1.37$\times 10^{-6}$, and the average number of neutrons per fission is 2.71, leading to a calculated neutron emission rate of 4.8 neutrons per second from this source. A scan over a wide range of thresholds was performed at a constant temperature of 19\degc, adjusting the pressure setpoint in 5psi increments, between 30 and 75~psia, providing a sample of 10 nuclear-recoil thresholds between 1.58~keV and 15.9~keV. The background rate was also measured at each pressure without the source present. The result (Figure \ref{neutronrate}) shows a decreasing trend in both the background and the source rate vs. nuclear recoil threshold. This is the expected result, as the chamber becomes increasingly insensitive to low-energy neutrons, both environmental and from the source, as the nuclear recoil threshold is increased.

\subsection{Response to gamma rays}
PICO bubble chambers with \ctfe\ target fluid have typically been operated at nuclear recoil thresholds above 3~keV where the probability of a gamma ray nucleating a bubble in the detector is $\sim 10^{-10}$ and the total expected background from gamma radiation is $<<1$ event; however, desire to explore low-mass WIMP search parameter space motivates operating the detectors at lower thresholds, where a detailed understanding of the gamma response in fluorocarbon bubble chambers is critical. A recent overall analysis of PICO gamma calibration data concludes that the probability of a gamma ray nucleating a bubble is not simply related to the nuclear recoil threshold. Varying the pressure and temperature setpoints of a bubble chamber can produce different gamma nucleation probabilities even for the same nuclear recoil threshold; increasing the operation temperature is observed to increase gamma sensitivity \cite{ERpaper}.

\begin{figure}[h!]
\centering
\includegraphics[scale=0.4]{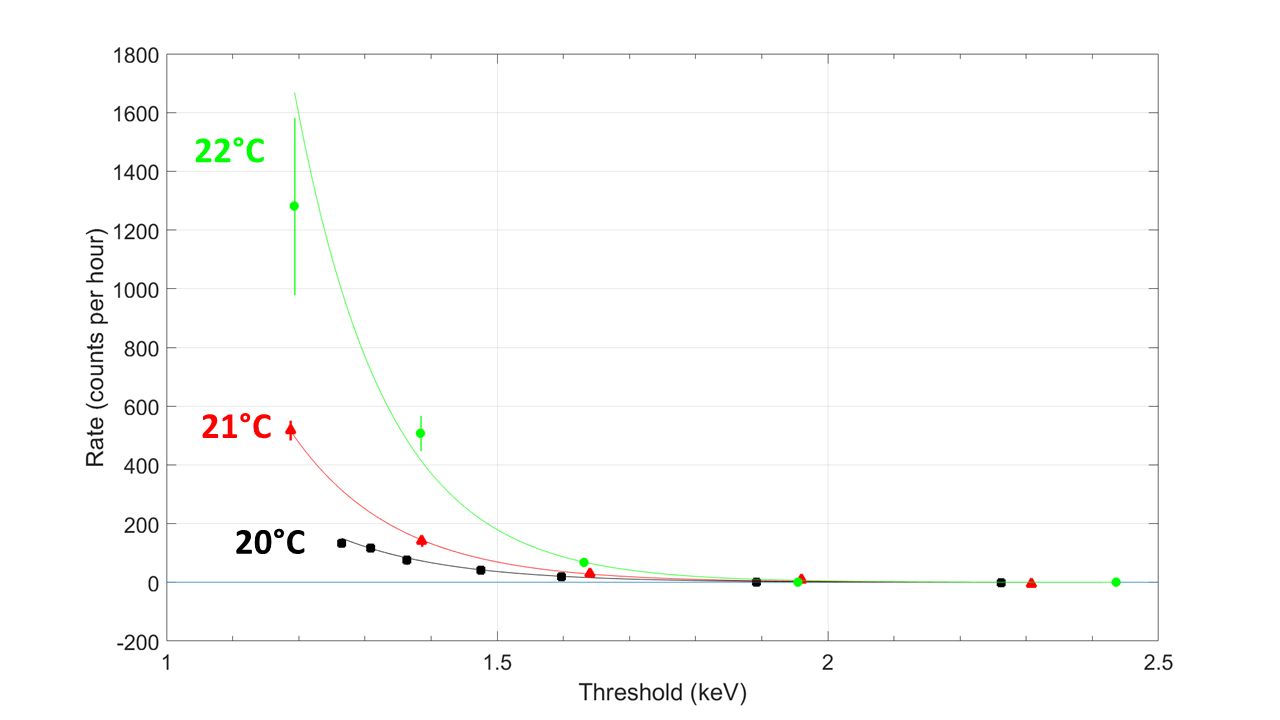}
\caption{\label{gammaturnon} Background-subtracted event rates in the DBC when operated with a gamma source present are shown with statistical error bars. The detector becomes sensitive to gamma rays between 1.5~keV and 2~keV nuclear recoil thresholds, when operated at 20\degc\ (black squares), 21\degc\ (red triangles), and 22\degc\ (green circles), with the 22\degc\ setting being the most sensitive to gammas at any given threshold. Exponential fits to the three data sets are shown as solid lines.}
\end{figure}

The DBC is sensitive to gamma rays when operated at low ($\lesssim 2$keV) nuclear recoil thresholds, corresponding to high temperature and low pressure. The threshold dependence of the gamma efficiency was measured at 20\degc, 21\degc, and 22\degc. The scans were performed with an 82~$\mu$Ci $^{137}$Cs source placed 25~cm away from the chamber, varying the pressure within each constant-temperature scan. The 20\degc\ scan consists of seven points and gives a sample of nuclear recoil thresholds between 1.26~keV and 2.29~keV, and the 21\degc\ and 22\degc\ scans are coarser, with 5 points each, over similar thresholds. The lowest nuclear-recoil threshold in the 21\degc\ scan is 1.19~keV, consistent with the lowest reported threshold of 1.2~keV at which the PICO-60 bubble chamber was operated\cite{60complete}. The results are illustrated in figure \ref{gammaturnon} and listed in table \ref{gammatable}. At thresholds above 1.75~keV, there is no significant rate above background, but at thresholds below 1.75~keV, a rate is observed, indicating that the DBC becomes sensitive to gamma rays near 1.7~keV. The temperature effect on gamma sensitivity is apparent by the variation of rate at constant threshold. 

\begin{center}
\begin{table}
\caption{\label{gammatable} The data from the gamma ray sensitivity scans. Mean nuclear recoil thresholds are calculated using the mean temperature and pressures recorded during the tests. Rates are given with statistical error bars after background subtraction, and are rounded to the nearest integer.}
\resizebox{\textwidth}{!}{%
\begin{tabular}{c|c|c|D{,}{\pm}{-1}}
\hline
Mean Nuclear Recoil Threshold (keV) & Temperature Set Point (\degc) & $P_{\text{chamber}}$ Set Point (psia) &  \multicolumn{1}{c}{Rate (counts per hour)}\\
\hline
\hline
2.44 & 22 & 55 &  1 , 7\\
\hline
1.96  & 22 & 50 &  1,5\\
\hline
1.63  & 22 & 45 & 68,7\\
\hline
1.38 & 22 & 40 & 508,60\\
\hline
1.19 & 22 & 35 & 1280,300\\
\hline
\hline
2.31 & 21 & 50 & -5,6\\
\hline
1.96 & 21 & 45 & 8,6\\
\hline
1.64 & 21 & 40 & 28,7\\
\hline
1.39 & 21 & 35 & 141,20\\
\hline
1.19 & 21 & 30 & 517,34\\
\hline
\hline
2.26 &20 & 45 &  -2,7\\
\hline
1.89 & 20 & 40 & 1,5\\
\hline
1.60 & 20 & 35 & 18,8\\
\hline
1.48 & 20 & 32.5 & 41,7\\
\hline
1.36 & 20 & 30 & 75,14\\
\hline
1.31 & 20 & 28.7 & 117,3\\
\hline
1.26 & 20 & 27.5 & 132,12\\
\hline
\end{tabular} }
\end{table}
\end{center}

\subsection{Acoustic Response}
\begin{figure}
\centering
\includegraphics[width = \textwidth]{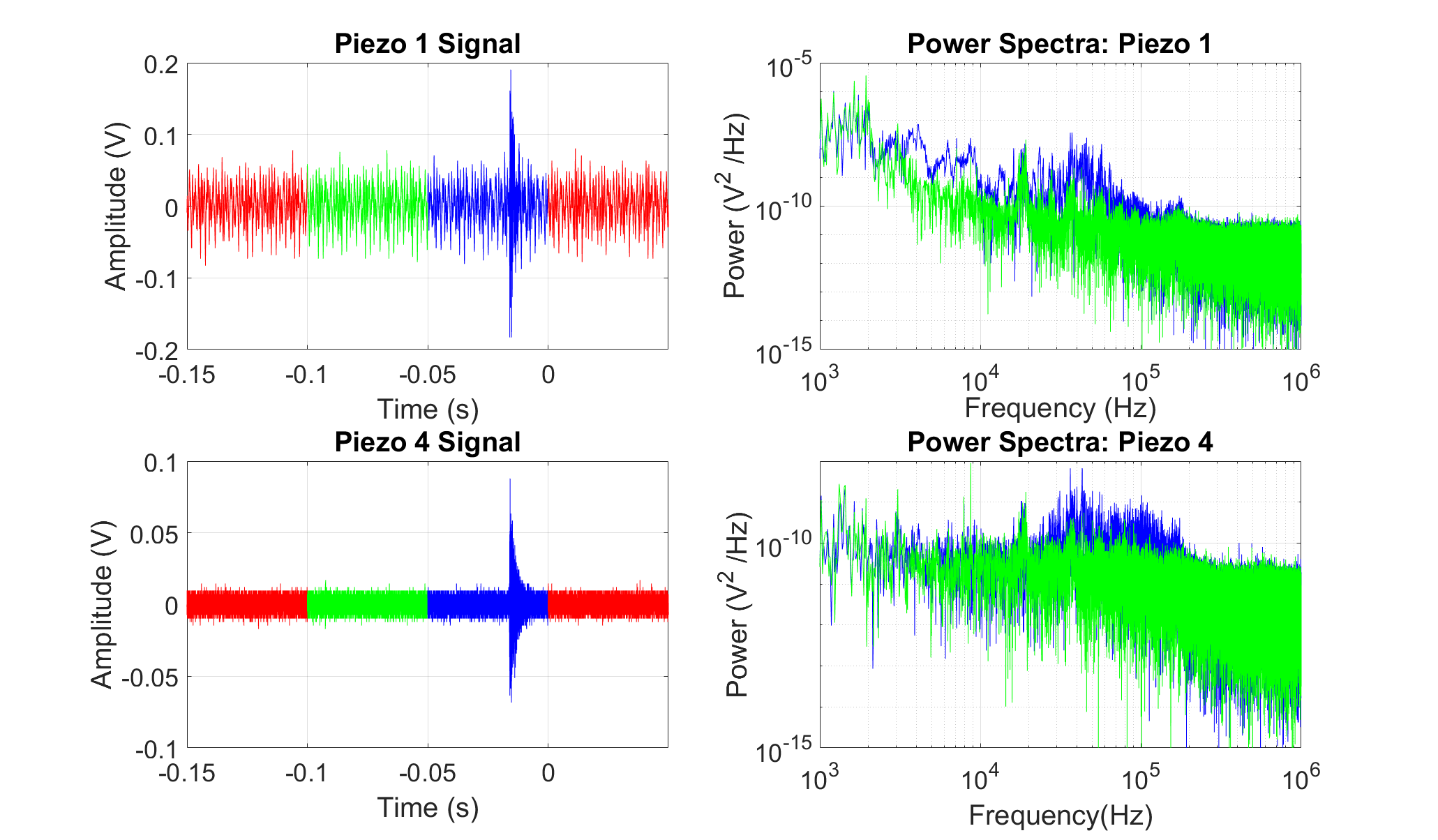}
\caption{\label{powerspectra} A typical event, shown in both piezos. Left: signals from the two piezos, the bubble occurs at approximately -0.015~s. The bubble's power is calculated in the blue part of the signal, the noise power is calculated in the green part, and the red part is not used in the acoustic analysis. The time at which the trigger comes in is considered $t=0$. The responses are not corrected for the gains or noise levels of the devices or amplifiers, which may vary. Right: power spectra shown for each piezo for both the pre-bubble noise and the bubble nucleation window. AP is interpreted as the area between the blue and green spectra in the range 55-120~kHz, which increases for louder bubbles whose sounds carry more power.}
\end{figure}

As with other bubble chambers, acoustic emission of bubbles in the DBC is quantified by the ``acoustic parameter'' (AP). The calculation of AP for the DBC follows a similar, although simplified, procedure to the AP calculation for PICO dark matter detectors. We begin by integrating the power spectrum of the acoustic trace between 55~kHz and 120~kHz to get the total power carried by the sound in that frequency range, which was the range used by PICO-60 for the AP calculation in \cite{pico60c3f8}. The power is calculated in both a pre-bubble window and the window where the bubble is nucleated, shown in the left panel of Figure \ref{powerspectra} for an example event. The right panel of Figure \ref{powerspectra} shows the difference between the power spectrum from the  noise and the bubble nucleation. Noise-subtraction is then performed and the parameter presented for piezo $i$ is 

\begin{equation}
AP_i=AP_{\text{bubble}} - AP_{\text{noise}}.
\end{equation}

\begin{figure}[h]
\centering
\includegraphics[width=\textwidth]{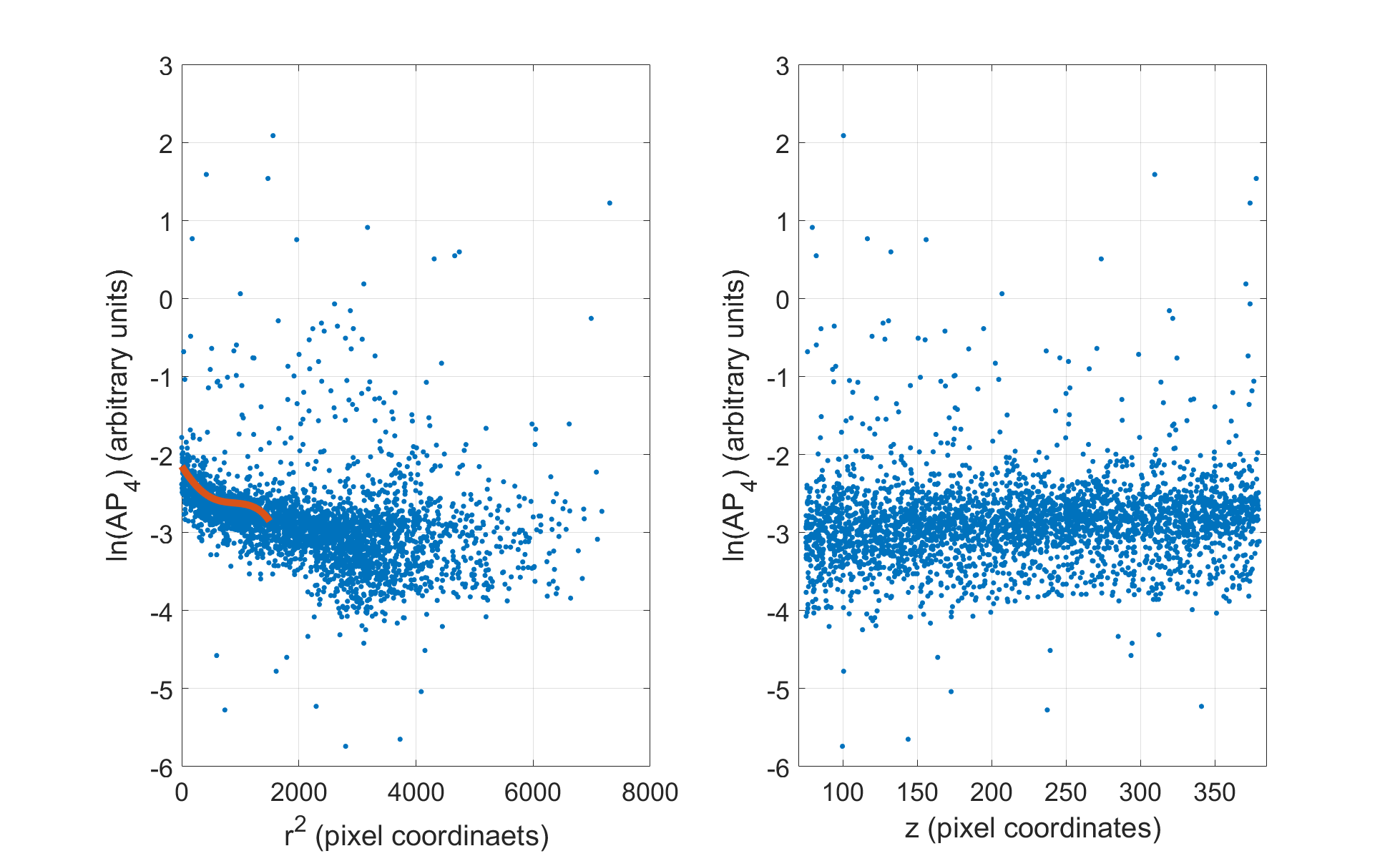}
\caption{\label{radAP} $\ln(AP_4)$ plotted against $r^2$ and $z$ for data taken at 30~psia; a cubic fit to the $r^2$ AP distribution shown as a solid line. Events near the center of the bubble chamber are about 2.5 times as loud as events near the wall, on average.}
\end{figure}

Bubbles nearer to the center of the bubble chamber are on average louder than those near the glass, though those bubbles nucleated very close to the glass itself do not fit this trend; there is a spread at the highest radii attributable to lensing and wall effects. We suspect that this radial dependence is the result of an interference effect between reflected and direct sound waves propagating in the fluid. The left frame of Figure \ref{radAP} data taken at $P_{\text{expand}}=30\text{~psia}$ to demonstrate this effect. This effect, and acoustic response overall, is best seen at lower pressures, because the signals from the piezos get louder with decreasing pressure.  Only a slight z dependence is observed, as demonstrated in the right frame of Figure \ref{radAP}, where the bubbles are seen to be slightly louder closer to the top of the chamber.

\begin{figure}
\centering
\includegraphics[scale=0.3]{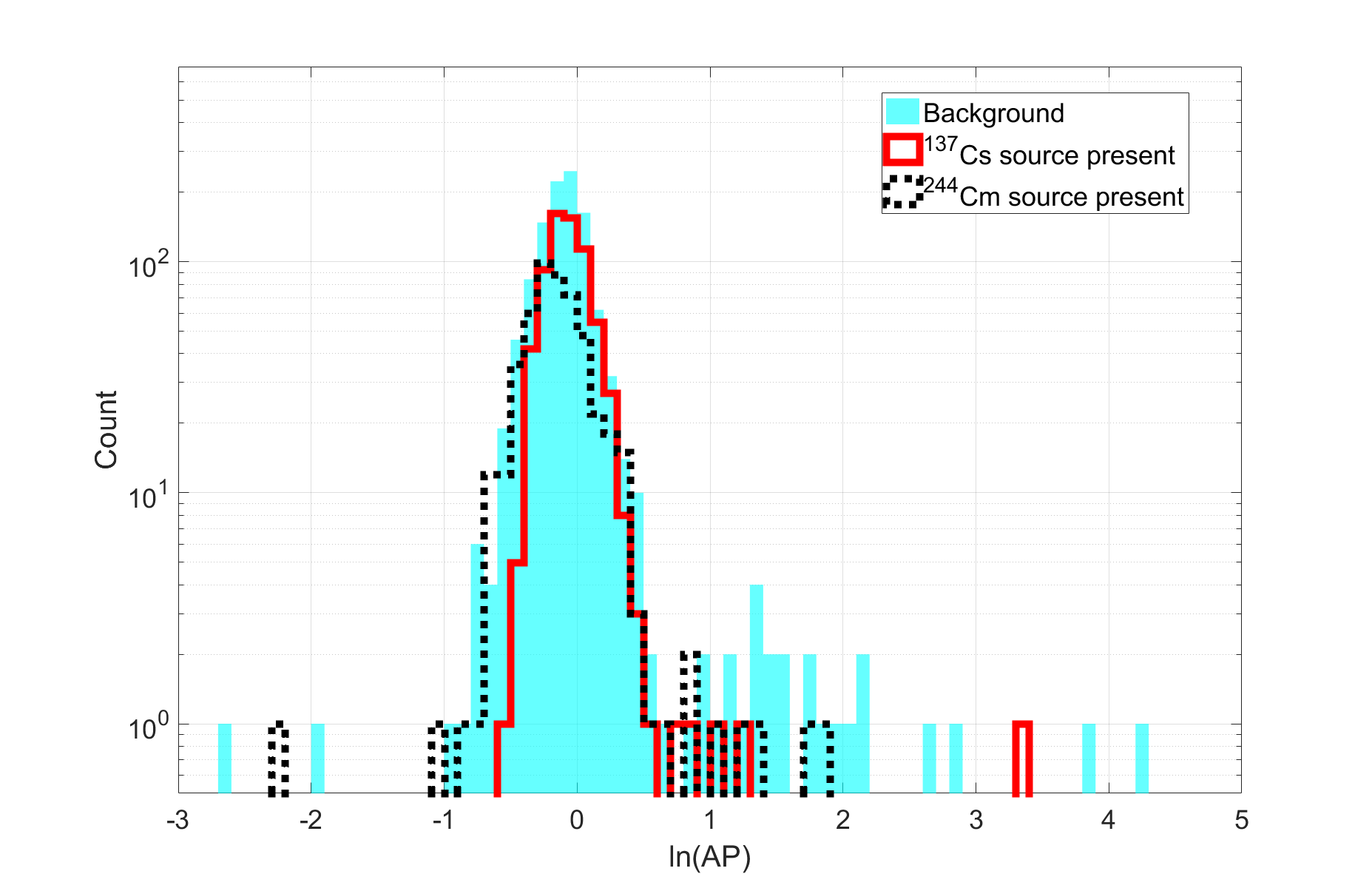}
\caption{\label{combinedAP} The measured AP distribution of three $P_{\text{expand}}=30\text{~psia}$, 20\degc\ runs. A background run is shown in shaded blue, a gamma source run with $^{137}$Cs is shown as a solid red line, and a neutron source run with $^{244}$Cm is shown as a dotted black line. High-AP events are consistent with high-energy neutron interactions and a background of alphas.}
\end{figure}

The radial dependence of the acoustic response is fit to a third-degree polynomial function for each piezo, for events in the central core of the chamber ($r^2 < 1500$, and $75<z<380$). Each value of $AP_i$ in that region is then corrected by dividing its value by the polynomial, normalizing the main band of the distribution to 1. The piezos' corrected responses $AP_{i\text{, corr.}}$ are then be combined into a single parameter by taking an equal-weight average, namely 
\begin{equation}
AP = \frac{AP_{\text{1, corr.}} + AP_{\text{4, corr.}}}{2}
\end{equation}
which is the parameter shown, after taking its natural logarithm, as a histogram in Figure \ref{combinedAP}. Increasing the degree of the polynomial fit was not observed to impact the result, and the fit function serves only as an empirical normalization method rather than a physical model. 

Figure \ref{combinedAP} shows the distribution of the combined parameter $\ln(AP)$ for a neutron source run, gamma source run and background run. The background is dominated by neutron scatter events, although a smaller population of alpha decay events is anticipated from dissolved radon. The runs were taken with identical thermodynamic conditions: $P_{\text{expand}}=30\text{~psia}$ and 20\degc, for a nuclear recoil threshold of 1.3keV. For all runs the bulk of the distribution is centered at $\ln(AP)=0$, as expected from previous PICO measurements showing no significant acoustic discrimination between nuclear recoils from neutrons and electron recoils from gamma rays. In all cases a small population of events extending to higher AP than the main peak is observed. This ``mid-AP tail'' has also been observed in previous PICO neutron calibration datasets. In the background run there is an additional excess of events in the $1.3 < \ln(AP) < 3.0$ region, relative to the source calibration runs. This is the region of expected alpha decay events, thus we identify the excess in this region as due to alpha-decay. Typically alpha decay events are observed in the $1 \lesssim \ln(AP) \lesssim 3$ region (see for example Figure 5 of \cite{60cf3i}), with the exact location varying based on alpha energy and detector specific factors. In this small chamber it was not possible to observe the timing correlations from radon chain alphas necessary to definitively identify radon decay alpha events. 

\begin{figure}
\centering
\includegraphics[scale=0.25]{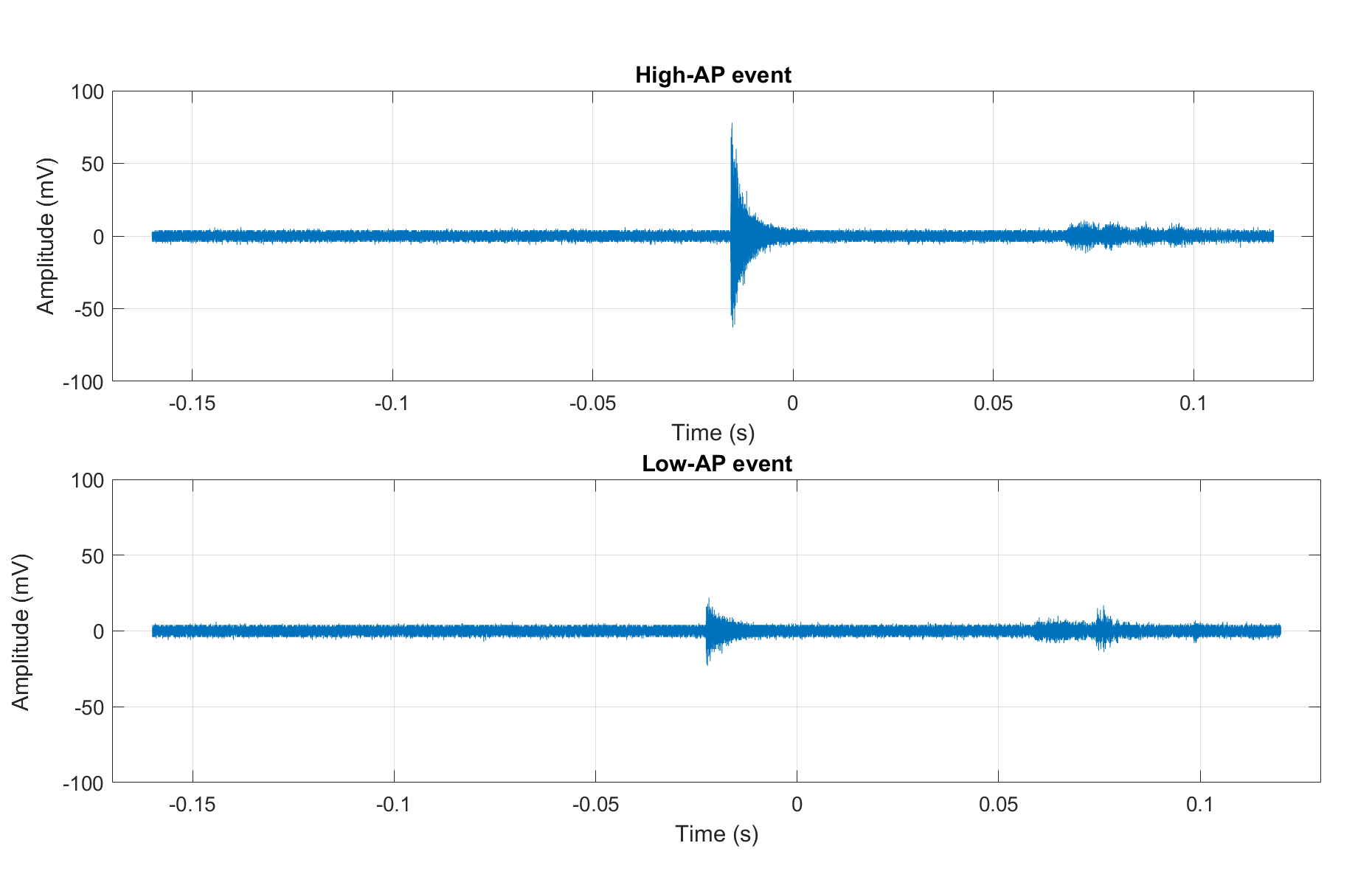}
\caption{\label{neutroncomparison} The responses from piezo 4 of events in the low-AP (neutron-like) and high-AP (alpha-like) regions are shown. The sound between -0.05~s and 0~s is from the bubble nucleation, whereas the sound between 0.05~s and 0.1~s is due to the compression of the chamber.}
\end{figure}

Event rates in the main peak, mid-AP tail ($0.7 < \ln(AP) < 1.3$) and ``high-AP alpha-like'' ($1.3<\ln(AP)<3.0$) regions are summarized in Table \ref{acousticstable}. The fraction of mid-AP events is similar between runs. The fraction of high-AP alpha-like events is higher in the background run than the source runs, as expected for the small contribution to the background from alpha decay of dissolved radon. However the rate per hour of high-AP events is somewhat higher in the neutron source run, suggesting the mid-AP tail of neutron events may extend into the high-AP alpha-like region. The exact nature of the highest AP neutron events is unknown, but they may include multiple bubble events with bubbles nucleated too close together to visually resolve and high-energy neutron-nucleus inelastic collisions. A comparison of the raw piezo traces of a neutron-like event and a high-AP alpha-like event is shown in Figure \ref{neutroncomparison}.

\begin{center}
\begin{table}
\caption{\label{acousticstable}A summary of the acoustic data, limited to the central core of the DBC.}
\resizebox{\textwidth}{!}{%
\begin{tabular}{||c|c|c|c|c||}
\hline
Data & Live Time & \begin{tabular}{@{}c@{}}Number of Single Bubbles \\ Number of Mid-AP Single Bubbles \\ Number of High-AP Single Bubbles\end{tabular} & Rate [Events per hour] & Fraction of Single Bubble Events  \\
\hline
Background & 121.1~hr & \begin{tabular}{@{}c@{}}1090\\7\\17\end{tabular} & \begin{tabular}{@{}c@{}}9.0$\pm$0.3\\0.06$\pm$0.02\\0.14$\pm$0.03\end{tabular} & \begin{tabular}{@{}c@{}}100\%\\ 0.6\% \\ 1.6\% \end{tabular}\\
\hline
$^{137}$Cs Gammas & 18.97~hr & \begin{tabular}{@{}c@{}}672\\4\\0 \end{tabular} & \begin{tabular}{@{}c@{}}35$\pm$1\\0.2$\pm$0.1\\0.0\end{tabular}  & \begin{tabular}{@{}c@{}}100\%\\ 0.6\%\\0\% \end{tabular}\\
\hline
$^{244}$Cm Neutrons & 9.26~hr & \begin{tabular}{@{}c@{}}499\\4\\3\end{tabular} & \begin{tabular}{@{}c@{}}54$\pm$2\\0.4$\pm$0.2\\0.3$\pm$0.2\end{tabular}  & \begin{tabular}{@{}c@{}}100\%\\0.8\%\\0.6\%\end{tabular}\\
\hline
\end{tabular} }
\end{table}
\end{center}

\subsection{Chamber Stability}\label{stability}
\begin{figure}[h!]
\centering
\includegraphics[width=\textwidth]{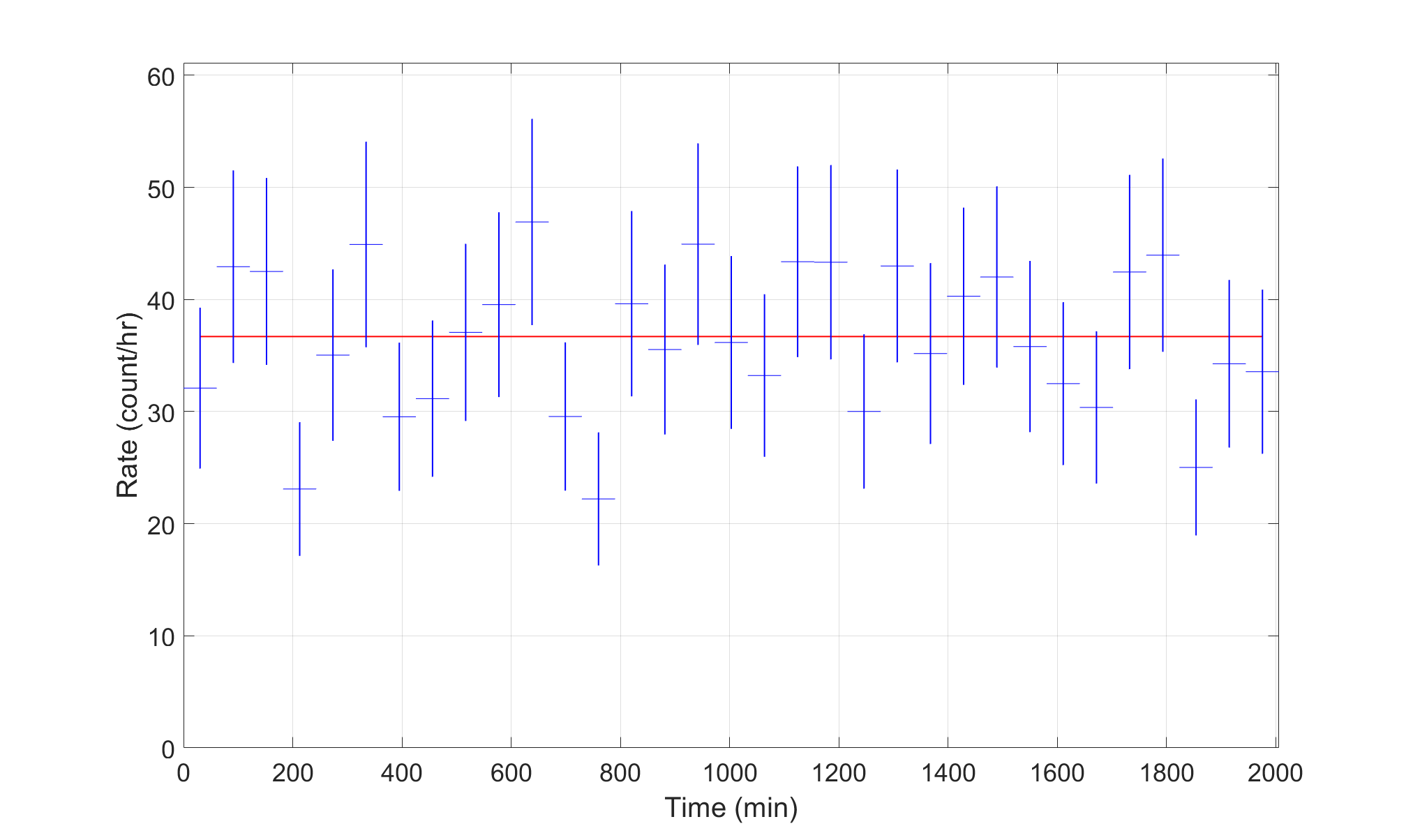}
\caption{\label{const_rate} Single-bubble rates during the 33.4 hour run, with cuts $z > 75$ and expand time $>3.14~\text{s}$. The run was divided into 33 evenly-sized bins, approximately one bin for each real hour of operation, and the rate was calculated within each time window. The average rate based on a constant fit to this binning is 36.7 events per hour, indicated by the solid red line, with $\chi^2/n.d.f.=23.7/32$.}
\end{figure}

A run was taken where the chamber was allowed to continuously operate over a period of 33.4 hours with no radioactive source present. The expansion pressure was $P_{\text{expand}}=35\text{~psia}$ and the chamber temperature was 20\degc, which gives a nominal nuclear recoil threshold of 1.5~keV. During this run, 728 single bubble events and 36 multiple-bubble events passing the z and event expand time cuts were recorded in 20.1 hours of detector live time (giving a live fraction of 60.2\%). This gives a mean overall rate of 38$\pm$1 events per hour, and a mean single-bubble rate of 36$\pm$1 events per hour. The rate is stable over the course of the run, as shown in Figure \ref{const_rate}. The total number of triggers during the run was 1111, so the mean trigger rate per live hour is 55$\pm$2, and the probability that a trigger resulted in a single bubble passing all cuts is $728/1111 = 65.5\%$. 

\begin{figure}[h]
\centering
\includegraphics[scale=0.2]{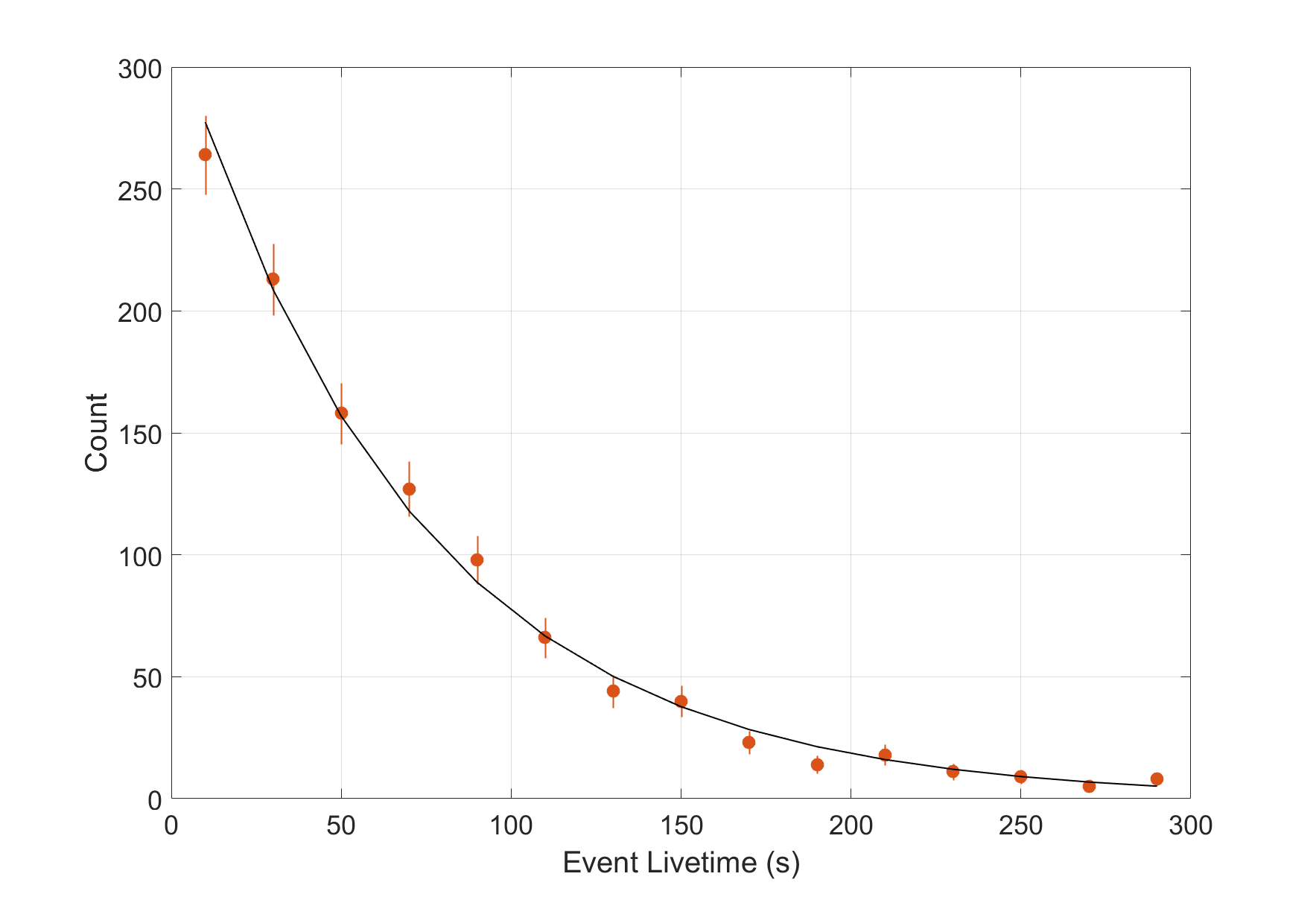}
\caption{\label{LT} Histogram of the live time of all the events in the run and exponential decay fit $y=319.88e^{-0.014255x}$, which excludes the last bin of the histogram that represents timeouts. Fit shows good agreement with exponential decay; $\chi^2/n.d.f. = 9.3/14$. }
\end{figure}

The low event rate means that the chamber can sometimes stay expanded for a few minutes before an event occurs, which leaves time to test the stability of the detector against spurious triggers unrelated to particle detection, such as nucleations on steel parts or particulate. The timeout was set to 300 seconds, and out of the 1111 events 16 were timeouts. About 90\% of the triggers were issued by the cameras, with the other 10\% made up of timeouts and pressure-related triggers. The fractions did not vary significantly in time. The live (expanded) time distribution for all events is shown in Figure \ref{LT}. The distribution of live time for all events shows good agreement with an exponential decay, as expected. From the fit, the trigger rate per live hour is 51.3~h$^{-1}$, which yields a single-bubble rate of 33.6~hr$^{-1}$ after multiplying by the the probability that a given trigger was a single bubble that passed all cuts.

As shown in Figure \ref{proj}, the DBC does not favor any region within the acceptance region after the basic cuts are made. This indicates that the detector is uniformly sensitive to the (mostly-neutron) background. Of particular note, there is no observed excess on the glass chamber walls. This is in contrast to previous PICO bubble chambers which have significant populations of events on the walls, which have been attributed to effects caused by the water in those chambers.

\subsection{Delayed Compression}
The future PICO-500 chamber will be large enough to detect multiple neutrinos from a nearby supernova through Coherent Elastic Neutrino-Nucleus Scattering (CE$\nu$NS), so can be operated as a supernova detector \cite{PICOSN}. It is expected that the detection in a one ton \ctfe\ bubble chamber of one neutrino from a nearby (10~kpc) Type-II supernova would be followed by another bubble within about 1.2 seconds and a third bubble within about 4 seconds. The ordinary method of compressing the chamber immediately upon detection of one bubble would not allow the capture of the subsequent events, but if the chamber can remain active for a few seconds after a bubble is observed, the chamber can be operated as both a WIMP search and a supernova detector.

\begin{figure}[h!]
\centering
\includegraphics[scale=0.4]{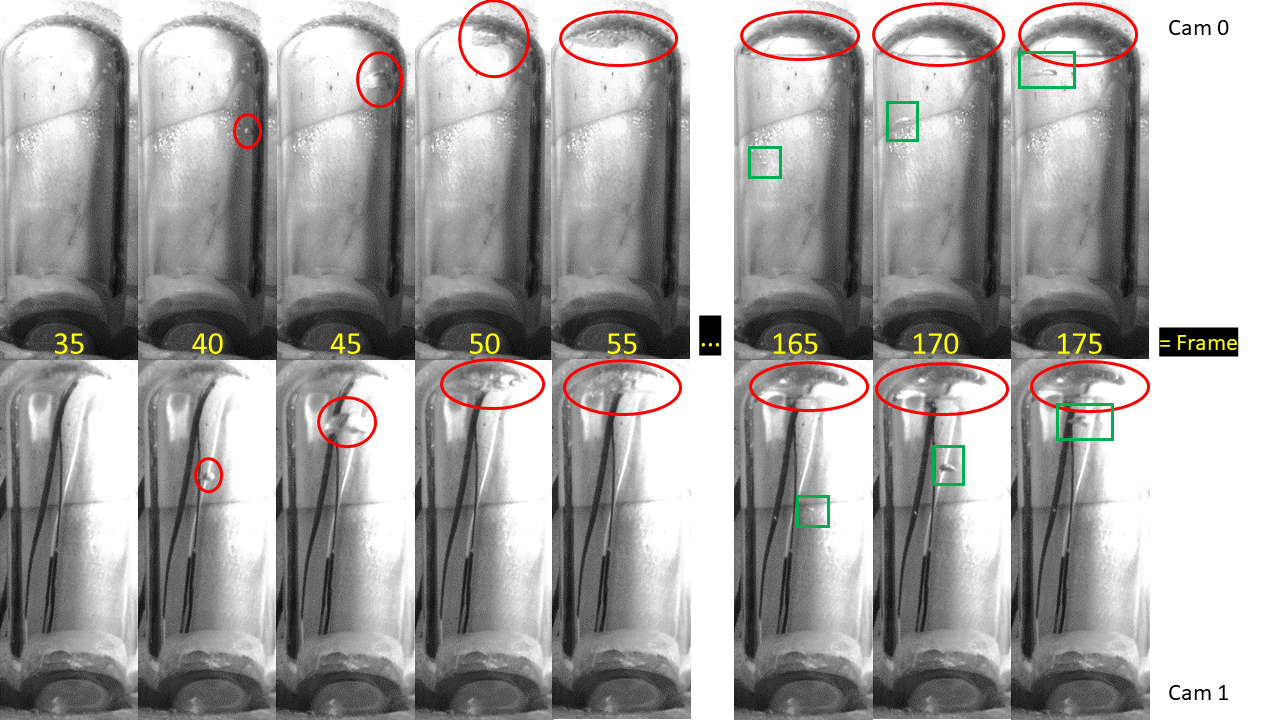}
\caption{\label{multiple_nucleation} Selected images from both cameras are shown from a multiple-nucleation event; frames are labeled in yellow. The first bubble is shown in red ovals, the second bubble is shown in green rectangles. Frames were collected at 50~Hz, so the second bubble was nucleated 2.5 seconds after the first.}
\end{figure}

As a proof of concept of delayed compression, the $^{244}$Cm source was used to produce multiple bubble nucleations within one expansion of the DBC. Triggering on the cameras and pressure transducers was disabled, and images were sampled at 50~Hz for 4 seconds at a time. This was done with entirely manual operation; the source was placed next to the chamber, the chamber was allowed to expand, and after a bubble was observed, a few seconds were allowed to pass before the system was triggered manually. The 200 images saved at each trigger were scanned manually for evidence of a multiple-nucleation event. An example of one such event is shown in Figure \ref{multiple_nucleation}, where two bubbles were nucleated 2.5 seconds apart. The nominal nuclear recoil threshold for the first nucleation of this event was 2~keV, however a pressure rise is observed after the first bubble as liquid continues to boil, so the threshold at the time of the second nucleation is not well constrained in a chamber as small as the DBC, where the pressure changes and cooling effects from the boiling propagate quickly through the entire active volume. This is not expected to be a problem in a detector as large as PICO-500, in which the amount of liquid boiled after 4 seconds is still very small compared to the fiducial mass.

\section{Conclusion}
The Drexel Bubble Chamber has successfully operated at thresholds at and below those used by PICO to search for WIMPs in the past, while maintaining the expected response to neutrons, gamma rays, and alpha decays. This RSU technology will improve PICO's dark matter search reach in PICO-40L and PICO-500 through its low-background and low-threshold design. We have demonstrated that the operation of a buffer-free bubble chamber is possible and practical over a wide range of nuclear-recoil thresholds, and such a bubble chamber can be made more or less sensitive to electron recoils depending on thermodynamic conditions. We have also conclusively demonstrated the ability of a bubble chamber to remain superheated for at least 2.5~s after the nucleation of one bubble; a capability significant for future uses of bubble chambers as underground supernova detectors.

\acknowledgments

We wish to acknowledge all members of the PICO collaboration, without whose scientific and technical developments this work would not have been possible. In particular we thank Scott Fallows, Guillaume Giroux and Hugh Lippincott for providing assistance in data transfer and management, and Ilan Levine, Edward Behnke, Haley Borsodi, and Thomas Nania at Indiana University South Bend for producing the piezoelectric acoustic transducers used in this work, with the support of National Science Foundation (NSF), Grant 1506377. We acknowledge computing and technical support from Fermi National Accelerator Laboratory under Contract No. DE412 AC02-07CH11359.

\bibliography{Drexel_Instrumentation_paper}

\end{document}